\documentclass[ba]{imsart}

\RequirePackage{amsthm,amsmath,amsfonts,amssymb}
\RequirePackage[authoryear]{natbib}
\RequirePackage[colorlinks,citecolor=blue,urlcolor=blue,backref=page,backref=page]{hyperref}
\RequirePackage{graphicx}

\pubyear{2026}
\volume{TBA}
\issue{TBA}
\firstpage{1}
\lastpage{30}
\startlocaldefs
\theoremstyle{plain}

\newtheorem{theorem}{Theorem}[section]
\newtheorem{lemma}[theorem]{Lemma}
\theoremstyle{definition}

\theoremstyle{remark}

\usepackage{multirow}
\usepackage{subcaption}
\usepackage{graphicx}
\usepackage[capitalise,nameinlink]{cleveref}
\usepackage{float}
\usepackage{pifont}
\newcommand{\xmark}{\ding{55}}
\usepackage[ruled]{algorithm2e}
\usepackage{mathabx}
\usepackage{tikz}
\usetikzlibrary{shapes.geometric,arrows,fit,matrix,positioning}
\tikzset
{
	treenode/.style = {rectangle, draw=black, align=center, minimum size=1cm},
	treeleaf/.style= {circle, draw=black, align=center, minimum size=1cm}
}
\allowdisplaybreaks

\floatstyle{ruled}
\providecommand{\algorithmname}{Algorithm}
\floatname{algorithm}{\protect\algorithmname}
\Crefname{algocf}{Algorithm}{Algorithms}

\newcommand{\mb}[1]{\mathbf{#1}}

\newcommand{\code}[1]{\texttt{#1}}

\mathchardef\mhyphen="2D

\newcommand{\bbT}{\mathbb{T}}
\newcommand{\bbM}{\mathbb{M}}
\newcommand{\bbI}{\mathbb{I}}

\def\trans{^\top}

\newcommand{\sN}{\mathcal{N}}

\newcommand{\sM}{\mathcal{M}}
\newcommand{\sD}{\mathcal{D}}

\newcommand{\sF}{\mathcal{F}}

\newcommand{\sG}{\mathcal{G}}

\newcommand{\vb}{\mathbf{b}}

\newcommand{\vr}{\mathbf{r}}

\newcommand{\vt}{\mathbf{t}}

\newcommand{\vx}{\mathbf{x}}
\newcommand{\vy}{\mathbf{y}}

\newcommand{\vphi}{\boldsymbol{\phi}}

\newcommand{\veta}{\boldsymbol{\eta}}

\newcommand{\vmu}{\boldsymbol{\mu}}

\newcommand{\mA}{\mathbf{A}}

\newcommand{\mD}{\mathbf{D}}

\newcommand{\mI}{\mathbf{I}}

\newcommand{\mT}{\mathbf{T}}

\newcommand{\mV}{\mathbf{V}}

\newcommand{\R}{\mathbb{R}}

\newcommand{\Prob}{\mathbb{P}}

\newtheorem{condition}[theorem]{Condition}
 
\newtheorem{corollary}{Corollary}

\usepackage{xr}
\makeatletter
\newcommand*{\addFileDependency}[1]{
	\typeout{(#1)}
	\@addtofilelist{#1}
	\IfFileExists{#1}{}{\typeout{No file #1.}}
}
\makeatother

\usepackage{comment}
\newcommand{\tbi}[1]{\textbf{\textit{#1}}}
\newcommand{\tb}[1]{\textbf{#1}}

\newcommand{\BlackBox}{\rule{1.5ex}{1.5ex}}  
\ifdefined\proof
    \renewenvironment{proof}{\par\noindent{\bf Proof\ }}{\hfill\BlackBox\\[2mm]}
\else
    \newenvironment{proof}{\par\noindent{\bf Proof\ }}{\hfill\BlackBox\\[2mm]}
\fi
\newtheorem{remark}[theorem]{Remark}

\definecolor{darkred}{RGB}{180,0,0}
\definecolor{darkblue}{RGB}{0,0,180}

\endlocaldefs

\begin{document}

\begin{frontmatter}
\title{Functional BART with Shape Priors: A Bayesian Tree Approach to Constrained Functional Regression}
\runtitle{Functional BART with Shape Priors}

\begin{aug}
\author[A]{\fnms{Jiahao}~\snm{Cao}\ead[label=e1]{jiahao.cao@uth.tmc.edu}\orcid{0009-0008-3850-178X}},
\author[B]{\fnms{Shiyuan}~\snm{He}\ead[label=e2]{heshiyuan@btbu.edu.cn}\orcid{0000-0003-2593-4566}}
\and
\author[C]{\fnms{Bohai}~\snm{Zhang}\ead[label=e3]{bohaizhang@bnbu.edu.cn}\orcid{0000-0002-0016-9813}}
\address[A]{Department of Biostatistics and Data Science, The University of Texas Health Science Center at Houston\printead[presep={,\ }]{e1}}

\address[B]{School of Mathematics and Statistics,
Beijing Technology and Business University\printead[presep={,\ }]{e2}}

\address[C]{Guangdong Provincial/Zhuhai Key Laboratory of IRADS,
Beijing Normal-Hong Kong Baptist University\printead[presep={,\ }]{e3}}

\runauthor{Cao et al.}
\end{aug}

\begin{abstract}
Motivated by the remarkable success of Bayesian additive regression trees (BART) in regression modeling, we propose a novel nonparametric Bayesian method, termed Functional BART (FBART), tailored specifically for function-on-scalar regression. FBART leverages spline-based representations for functional responses coupled with a flexible tree-based partitioning structure, effectively capturing complex and heterogeneous relationships between response curves and scalar predictors. To facilitate efficient posterior inference, we develop a customized Bayesian backfitting algorithm. Additionally, we extend FBART by introducing shape constraints (e.g., monotonicity or convexity) on the response curves, enabling enhanced estimation and prediction when prior shape information is available. The use of shape priors ensures that posterior samples respect the specified functional constraints.  Under mild regularity conditions, we establish posterior convergence rates for both FBART and its shape-constrained variant, demonstrating rate adaptivity to unknown smoothness. Extensive simulation studies and analyses of two real datasets illustrate the superior estimation accuracy and predictive performance of our proposed methods compared to existing state-of-the-art alternatives.
\end{abstract}

\begin{keyword}[class=MSC]
\kwd[Primary ]{62F15}
\kwd{62G05}
\kwd[; secondary ]{62R10}
\end{keyword}

\begin{keyword}
\kwd{BART}
\kwd{Bayesian nonparametrics}
\kwd{function-on-scalar regression}
\kwd{posterior concentration}
\kwd{shape constrained inference}
\end{keyword}

\end{frontmatter}

\section{Introduction}
The increasing availability of complex and high-resolution data has brought functional data analysis  \citep[FDA; see][]{ramsay1991some,ramsay2005functional,wang2016functional} to the forefront of modern statistical methodology. The FDA method often leverages  intrinsic data structures such as smoothness to address high dimensionality challenges and enhance estimation efficiency.  Functional responses—such as curves or surfaces—naturally arise in a wide range of regression applications, including growth curve modeling across diverse domains~\citep{tang2008pairwise,severson2019data,fan2022conditional}, neuroimaging studies of critical diseases~\citep{zhang2022high,zhu2023statistical}, and the modeling of yield and Lorenz curves in economic and financial analyses~\citep{hays2012functional,jann2016estimating,kowal2019functional}. In these datasets, response curves often exhibit complex and nonlinear relationships with covariates and, in many cases, are subject to known structural constraints such as monotonicity or convexity. For instance, in economics, the call price of a European option must be both decreasing and convex in the strike price \citep{birke2007estimating}, while wage profiles are typically expected to be concave in years of work experience \citep{hannah2013multivariate}. Accurate modeling in such settings requires methods that are not only flexible and robust but also capable of incorporating prior knowledge about the functional shape to improve estimation efficiency and interpretability~\citep{groeneboom2014nonparametric,horowitz2017nonparametric,ghosal2023shape}.

{In the functional regression literature, either the response, the covariates, or both may be functions \citep{chiou2004functional,yao2005functional,morris2015functional,greven2017general,  he2023unified}. This work focuses on the function-on-scalar regression (FOSR) setting, where the response is a function and the predictors are scalars. Classical FOSR methodologies, particularly functional linear models, have been widely adopted due to their interpretability and theoretical tractability \citep{morris2006wavelet, rosen2009bayesian, morris2015functional, chen2016variable, kowal2020bayesian, ghosal2023shape}. However, their linearity assumption can be restrictive when the underlying relationship exhibits nonlinearities or complex covariate interactions. In parallel, substantial effort has been devoted to modeling and analyzing the shape of functional data, motivated by domain-specific requirements and interpretability considerations. One prominent paradigm uses shape-constrained regression splines, imposing properties such as monotonicity or convexity through restrictions on spline coefficients \citep{ramsay1998estimating,meyer2008inference, abraham2015bayesian, pya2015shape, wang2025monotone}. Another direction focuses on the intrinsic geometry of functional curves by developing transformation-invariant representations, elastic metrics, and size--shape decompositions \citep{marron2015functional, srivastava2016functional, wu2024shape, kundu2025decomposition}.

 Despite these developments, flexible nonlinear FOSR methods that can simultaneously accommodate complex covariate effects and functional shape constraints remain relatively limited. \citet{scheipl2015functional} introduced a broad additive regression framework capable of capturing linear and nonlinear effects across scalar and functional predictors via tensor-product splines over covariates $\mathbf{x}$ and functional index $t$. Within the Fr\'echet regression paradigm \citep{petersen2019frechet}, \citet{fan2022conditional} developed global and local functional regression estimators; however, their local kernel-smoothing approach is primarily restricted to low-dimensional continuous covariates. \citet{kundu2025decomposition} proposed a shape-constrained Fr\'echet regression model for positive and monotone functional responses by decomposing each function into distinct shape and size components, each equipped with its own metric. While these methodologies represent significant progress, there remains a need for flexible nonlinear FOSR frameworks, particularly within a Bayesian framework, which can seamlessly enforce shape constraints while providing coherent uncertainty quantification through posterior distributions.}

 Our approach is motivated by the remarkable success of Bayesian additive regression trees (BART) in a variety of regression settings \citep{chipman2010bart, hill2020bayesian}. The BART model, as an ensemble of multiple Bayesian regression trees \citep{chipman1998bayesian, denison1998bayesian}, has gained popularity due to its inherent flexibility, strong predictive performance, and natural capacity for uncertainty quantification. Recent developments have significantly expanded BART's applicability, with advances in domain partitioning strategies \citep{ge2019random, luo2021bast,he2025gs}, dimension reduction capacity and smoothness adaptation \citep{linero2018bayesianHD, linero2018bayesian, rovckova2020posterior, liu2021variable}, formal inferential procedure \citep{castillo2021uncertainty}, {extensions beyond piecewise-constant leaf models through local linear regressions \citep{prado2021bayesian, herren2025stochtree} and basis-function expansions \citep{filipini2025tree},} and  complex  data handling \citep{li2023adaptive, um2023bayesian,diaz2025nonparametric}. Yet, existing BART framework focuses on scalar outputs, and cannot naturally and efficiently process functional responses by exploiting their intrinsic smoothness property.
 
In this work, we introduce a fully nonparametric Bayesian tree model for the FOSR problem, termed \textit{Functional Bayesian Additive Regression Trees (FBART)}.
 Our proposed model advances the FOSR literature as well as the BART literature: By combining spline-based function representations with tree-based domain partitioning, FBART can effectively model functional responses and capture highly nonlinear relationships. To further improve interpretability and incorporate domain knowledge, we develop a shape-constrained version of FBART, referred to as \textit{S-FBART}, and provide a corresponding inference procedure. In particular, we employ a basis representation approach for modeling shape-constrained functions 
 where, for appropriately chosen basis functions, shape constraints on real-valued functions can be enforced through a set of linear constraints on the basis coefficients. To facilitate
efficient posterior inference, we develop tailored Bayesian backfitting algorithms for both FBART and its shape constrained variant. Comprehensive numerical studies show that the proposed algorithms work very well. 

From a theoretical perspective, we establish posterior contraction rates for both FBART and S-FBART under mild regularity conditions. 
To the best of our knowledge, theoretical results on Bayesian tree-based methods for function-on-scalar regression—particularly with shape constraints—have not been previously explored in the literature.  Establishing these theoretical properties presents substantial challenges; we overcome these by constructing novel sieve spaces and designing suitable regularizing priors that account for the joint complexity introduced by both spline basis dimension and tree-based domain partition structures. By carefully leveraging spline approximation theory and Bayesian tree priors, our models (FBART and S-FBART) achieve an effective balance between estimation bias and variance. Specifically, the proposed models maintain appropriate model complexity and desirable prior mass concentration near good approximations without requiring explicit knowledge of smoothness parameters. 

In the literature, two lines of research that are closely related to this work have recently emerged. The first is BART with targeted smoothing \citep{starling2019monotone, starling2020bart}, which induces smooth variation over a specified covariate by placing Gaussian process priors on the terminal nodes of trees, and imposing monotonicity through posterior projection \citep{lin2014bayesian}. {\cite{filipini2025tree} further proposed replacing the Gaussian process priors in tsBART with basis-function expansions to improve computational scalability, while their approach does not accommodate shape constraints.} The second is the monotone BART model \citep{chipman2022mbart}, which ensures that the scalar response is monotonic in certain predictors. In contrast, our proposed FBART is a fully Bayesian approach explicitly designed for function-on-scalar regression, employing a flexible yet efficient spline-based representation. By directly leveraging the functional structure of the response, FBART achieves superior estimation accuracy and improved uncertainty quantification compared to existing methods, as demonstrated through simulation studies and real-data applications. The shape-constrained extension, S-FBART, naturally accommodates a diverse range of complex constraints—including but not limited to monotonicity—within a coherent Bayesian framework. Finally, we also provide theoretical guarantees for both FBART and S-FBART under the function-on-scalar regression framework, addressing a significant gap in prior research.

The rest of the paper is organized as follows. In \cref{sec:FBART,subsec:shapeFBART}, we introduce the FBART model and its shape-constrained version, respectively. In \cref{sec:convergence}, posterior concentration results of the proposed methods are established. In \cref{sec:simu}, we conduct simulation studies to demonstrate the performance of FBART and S-FBART, with comparisons to other competing FOSR methods;  the proposed methods are then applied to two real datasets in \cref{sec:RD}. Finally, a brief summary and discussion are given in \cref{sec:discussion}. Technical details and additional numerical results are given in the Supplementary Materials \citep{Cao2026functional}.

\section{Methodology}\label{sec:FBART}

\subsection{Notation and model setup}
We first introduce the mathematical notations used in this paper. Let $\|\cdot\|_q$ denote the $q$-norm of vectors and matrices, for $q \in [1, \infty]$. For a positive integer $j$, we use $[j]$ to denote
the set of consecutive integers $\{1,\ldots,j\}$. For a vector $\vb$, we use $\vb(i)$ to represent its $i$th entry. For a matrix $\mA$, $\mA(i, j)$ denotes its $(i, j)$th element. We use $\mb{0}$ to denote the zero vector and $\mathbf{I}_n$ to denote the identity matrix of size $n$. We use $\sN(\cdot, \cdot)$ to denote a (multivariate) normal distribution, and $\mathcal{N}(\cdot;\cdot, \cdot)$ to denote the corresponding density function. We use $\pi$ or $\pi_n$ to denote the prior distribution, and $\Pi_n$ for the posterior distribution. Given a set $A$, $\bbI_{A}(\cdot)$ denotes the indicator function on $A$.

Let $\mathcal{F}$ be the space of functions mapping from $\mathbb{R}$ to $\mathbb{R}$, which satisfy certain smoothness and (or) shape constraint. Suppose that for each subject $i = 1, \ldots, n$, we observe a functional response $Y_i\in \mathcal{F}$ along with a covariate vector $\vx_i \in \mathbb{R}^p$. Let $\Xi_0(\cdot) = \mathbb{E}(Y \mid \cdot)$ denote the true regression map from the covariate space $\mathbb{R}^p$ to the function space $\mathcal{F}$. We consider the following function-on-scalar regression model:
\begin{equation}\label{eq:fos}
    Y_i = \Xi_0(\vx_i) + \epsilon_i,\quad (i=1,\ldots,n),
\end{equation}
where $\epsilon_i$ is an independent Gaussian white noise process on $\R$ with variance $\sigma^2 \in \mathbb{R}^+$. Without loss of generality, we assume that the domain of the response functions is $[0, 1]$, and the covariate space is $[0, 1]^p$. In practice, each functional response $Y_i$ is observed at a set of $m_i$ points $\vt_i = \{t_{i1}, \ldots, t_{im_i}\} \subseteq [0, 1]$. The goal is to estimate the true regression map $\Xi_0$ based on the observed data $\big\{ (\vx_i, \{Y_i(t_{ij})\}_{j=1}^{m_i}) \big\}_{i=1}^n$.   

{The formulation in Equation~\eqref{eq:fos} extends functional linear models by allowing the regression map to depend nonlinearly on the scalar covariates. Defining $\Xi(t;\vx):=\Xi(\vx)(t)$, the model can equivalently be written pointwise as $Y_i(t_{ij})=\Xi_0(t_{ij};\vx_i)+\epsilon_i(t_{ij})$, where $\epsilon_i(t_{ij})\overset{\mathrm{i.i.d.}}{\sim}\sN(0,\sigma^2)$, making explicit the dependence of the conditional mean on both the functional index $t_{ij}$ and the covariates $\vx_i$. This pointwise representation is closely related to the targeted-smoothing framework of \citet{starling2020bart}, where the conditional mean is assumed to vary smoothly with respect to the target variable $t$. We retain the function-valued notation in Equation~\eqref{eq:fos} to emphasize that the regression target is the entire conditional mean function $\Xi_0(\vx)\in\sF$, whose structure over $t$ is represented through a B-spline expansion that naturally accommodates smoothness and shape constraints.}

\subsection{Review of Bayesian additive regression trees}\label{subsec:reviewBART}
We first briefly review the Bayesian additive regression trees \citep[BART,][]{chipman2010bart}, which model scalar-valued response with vector input. Overall, the BART model consists of two components: a sum-of-trees model and a regularization prior. 

\begin{figure}
    \centering
    \begin{subfigure}{0.5\textwidth}
        \centering
        \begin{tikzpicture}[->,>=stealth',level/.style={sibling distance = 8cm/#1,
					level distance = 1.5cm},scale=0.6, transform shape]
				\node [treenode] {$\vx(1)\leq0.8$}
				child
				{
					node [treenode] {$\vx(2)\leq0.5$}
					child
					{
						node [treenode] {$\vx(1)\leq0.3$} 
						child{	node [treeleaf] {$1$} }
						child{	node [treeleaf] {$2$} }
					}
					child
					{
						node [treenode] {$\vx(2)\leq0.7$} 
						child{	node [treeleaf] {$3$} }
						child{	node [treeleaf] {$4$} }
					}
				}
				child
				{
					node [treeleaf] {$5$} 
				}
				;
		\end{tikzpicture}	 \label{fig:chap3:treedemo}
    \end{subfigure}%
    \begin{subfigure}{0.5\textwidth}
        \centering\includegraphics[width=0.65\textwidth]{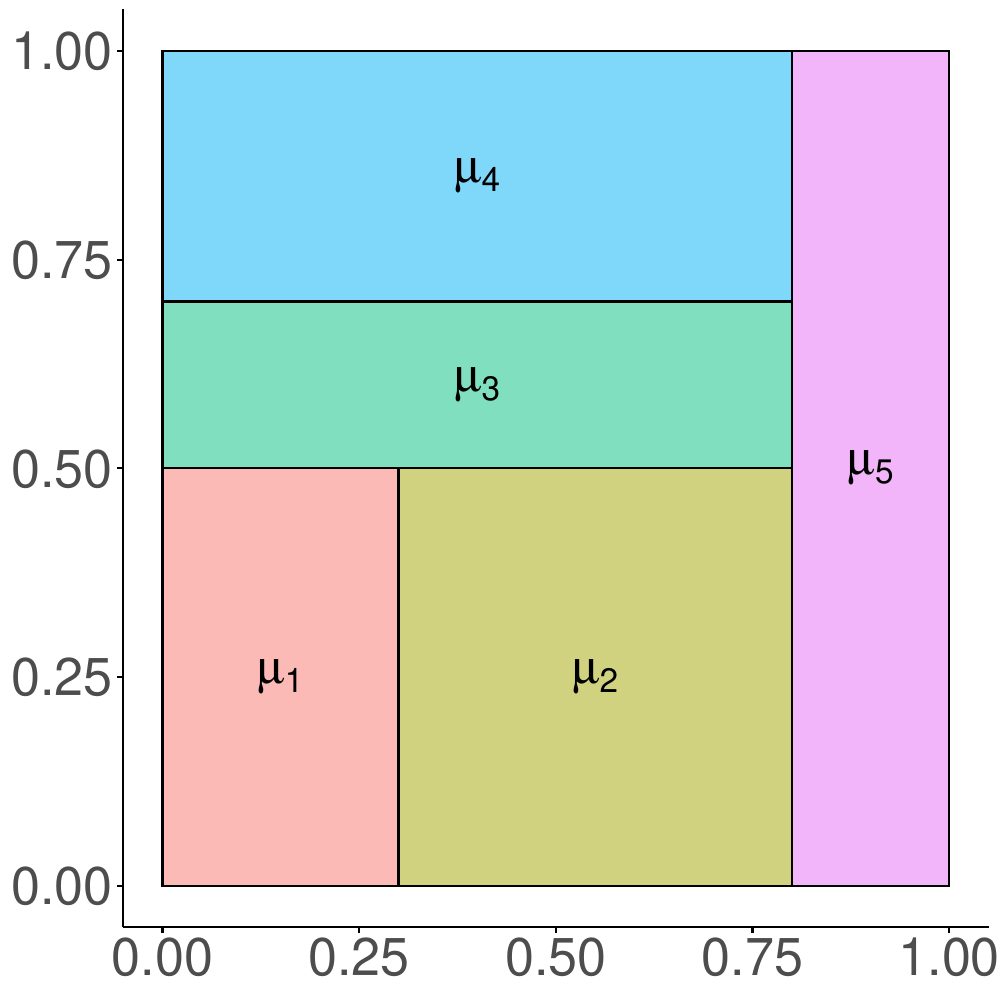}
        \label{fig:chap3:treepartitiondemo}
    \end{subfigure}
    \caption{A binary decision tree $\mT$ on $[0, 1]^2$ with $5$ terminal nodes (left panel), and a regression tree function $g(\cdot;\mT,\sM)$ with $\sM = \{\mu_{\ell}\}_{\ell=1}^5$ (right panel).}
    \label{fig:tree}
\end{figure}

As an ensemble Bayesian method, BART approximates a real-valued function $f(\cdot)$ on $\R^p$ by a sum of $K$ regression trees, denoted as $\sum_{k=1}^K g(\cdot;\mT_k,\sM_k)$, where each $g(\cdot;\mT_k,\sM_k)$ is a function parameterized by a binary decision tree $\mT_k$ and its associated terminal node parameters $\sM_k$. Specifically, a binary decision tree $\mT_k$ with $L_k$ terminal nodes (leaves) can be represented by a binary tree topology and a set of splitting rules for the internal nodes. The splitting rules are binary splits of the form $\{\vx:\vx(j)\leq z\}$ versus $\{\vx:\vx(j)> z\}$, where $\vx(j)$ is the splitting variable with $j\in[p]$ and $z\in\{\vx_i(j)\}_{i=1}^n$ is the splitting value selected from the observed values of the splitting variable. The terminal nodes of $\mT_k$ then yield a rectangular-shaped partition $\sD_k = \{D_k^1,\ldots,D_k^{L_k}\}$ of the covariate space. Given the node parameters $\sM_k= \{\mu_{k1},\ldots,\mu_{kL_k}\}\subseteq \R$, the $k$th regression tree function $g(\cdot;\mT_k,\sM_k) = \sum_{\ell=1}^{L_k} \mu_{k\ell}\times \mathbb{I}_{D_k^{\ell}}(\cdot)$ is  piece-wise constant. Figure~\ref{fig:tree} provides an illustrative example of a binary decision tree and its induced regression tree function.

To avoid overfitting, a regularization prior is imposed on the model parameters. In particular, the prior takes the form $\pi\big(\{\mT_k,\sM_k\}_{k=1}^K\big) = \prod_{k=1}^K\pi( \sM_k\mid\mT_k)\pi(\mT_k)$. 
For the node parameters $\sM_k$, conjugate normal priors are typically used to enable Gibbs sampling. 
The binary decision tree prior $\pi(\mT_k)$ is implicitly specified by the following tree-generating stochastic process. First, $\mT_k$ is initialized with a single root node with depth $d = 0$; the probability that a node at depth $d\geq 0$ splits (i.e., it is internal) is $p_{\text{split}}(d)$. For any internal node, its splitting rule is assigned by first sampling a splitting variable index $j$ uniformly from the available indices in $[p]$, and then sampling a splitting value $z$ uniformly from the available covariate values of the variable $\vx(j)$.  The splitting probability in \cite{chipman1998bayesian,chipman2010bart} takes the form $p_{\mathrm{split}}(d) = a_{\mathrm{split}}(1+d)^{-b_{\mathrm{split}}}$, where $a_{\mathrm{split}}\in(0,1)$ and $b_{\mathrm{split}}\geq 0$ are hyperparameters. Apparently, this prior penalizes the splitting probabilities for nodes of large depths.


\subsection{Functional BART via B-spline representation}\label{subsec:FBART}


We now extend the classical BART from modeling \textit{real-valued} responses to  \textit{function-valued} responses. For this purpose, we introduce a family of tree-structured maps from $[0,1]^p$ to $L_2([0,1])$, termed \textit{ functional regression tree maps}. The mapping is constructed with the B-splines, which stands out among the various basis representations for functional data due to its appealing theoretical properties and numerical advantages \citep{deboor1978practical, unser1993b}. 


The order-$q$ B-spline basis \citep{deboor1978practical} can be recursively defined as follows. Let $\{\xi_j\}_{j=1}^{J+q}$ be a knot sequence such that $\xi_{j+1} = \xi_{j}$ if $j\leq q-1$ or $j\geq J+1$, and $\xi_{j+1} > \xi_{j}$ otherwise. For $\underline{q}\leq q$, the B-spline basis functions $\{\phi_{j,\underline{q}}\}_{j=1}^{J+q-\underline{q}}$ of order $\underline{q}$ take the following form: 
\begin{equation*}\label{eq:Bspline}	
    \phi_{j,\underline{q}}(t) = \left\{
    \begin{aligned}
        & \mathbb{I}_{[\xi_j,\xi_{j+1})}(t),\quad \quad\underline{q}=1,\\
        &\frac{t-\xi_j}{\xi_{j+\underline{q}-1} - \xi_j}\phi_{j,\underline{q}-1}(t) + \frac{\xi_{j+\underline{q}}-t}{\xi_{j+\underline{q}} - \xi_{j+1}}\phi_{j+1,\underline{q}-1}(t),\quad\underline{q}>1.
    \end{aligned}\right.
\end{equation*}
For simplicity, we may omit the order $q$ in the subscript when no confusion arises, and denote by $\{\phi_j\}_{j\in[J]}$ a set of order-$q$ B-spline basis functions with boundary knots $\xi_1 = 0$ and $\xi_{J+q} = 1$.

Given a binary decision tree $\mT$ with $L$ leaf nodes and node parameters $\sM = \{\vmu_1,\ldots,\vmu_L\}\subseteq \R^J$, we refer to the following map $\Xi_{\mT,\sM}:[0,1]^p\to L_2([0,1])$ as a functional regression tree map:
\begin{equation*}
    \Xi_{\mT,\sM}(\cdot) = \sum_{\ell=1}^L \vphi\trans\vmu_{\ell}\times \mathbb{I}_{D^{\ell}}(\cdot) =  \sum_{\ell=1}^L   \Big\{\sum_{j=1}^J\phi_j\vmu_{\ell}(j)\Big\}\times \mathbb{I}_{D^{\ell}}(\cdot) ,
\end{equation*}
where $\vphi = (\phi_{1},\ldots,\phi_{J})\trans$ is the basis-function vector and $\sD = \{D^1,\ldots,D^{L}\}$ is the partition of $[0,1]^p$ induced by $\mT$. 

Next, we define the functional additive regression tree map as follows. Let $\{\mT_k\}_{k=1}^K$ denote a collection of $K\geq 1$ binary decision trees. For each $\mT_k$ with $L_k$ leaf nodes, the induced partition is $\sD_k = \{D_k^{\ell}\}_{\ell=1}^{L_k}$. Let $\sM_k = \{\vmu_{k\ell}\}_{\ell=1}^{L_k}\subseteq\R^J$ be the node parameters associated with $\mT_k$. By writing $\bbT = \{\mT_k\}_{k=1}^K$ and $\bbM = \{\sM_k\}_{k=1}^K$, the functional additive regression tree map is:
\begin{equation} \label{eqn:treemapdef}
    \Xi_{\bbT,\bbM}(\cdot) = \sum_{k=1}^K \Xi_{\mT_k,\sM_k}(\cdot) = \sum_{k=1}^K\sum_{\ell=1}^{L_k} \vphi\trans\vmu_{k\ell}\times \mathbb{I}_{D_k^{\ell}}(\cdot).
\end{equation}

Although we focus on the axis-aligned partition induced by binary decision trees in this paper, the above treatment is generic and other space partitioning methods can be incorporated. Possible alternatives include random tessellation forests \citep{ge2019random} and random spanning trees \citep{luo2021bast,luo2024nonstationary}.



\subsection{Prior specification and posterior inference}\label{subsec:prior_and_mcmc}

The functional additive regression tree map $\Xi_{\bbT,\bbM}$ in~\eqref{eqn:treemapdef} involves
 $K$ binary decision trees $\{\mT_k\}_{k=1}^K$ and their associated node parameters $\{\sM_k\}_{k=1}^K$. To complete the Bayesian model specification, we assign prior distributions to $\{\mT_k,\sM_k\}_{k=1}^K$ as well as to the noise variance $\sigma^2$. A possible extension is to treat $J$ and $K$ as unknown parameters and place discrete priors on them, estimating these quantities using a Metropolis-Hastings algorithm with random walk proposals. However, this approach can lead to considerable computational overhead. Following standard practice in the BART literature \citep[e.g.,][]{chipman2010bart}, we instead fix these integer parameters and provide default guidelines for their selection. In practice, they can be chosen via cross validation or other model selection criteria such as WAIC \citep{watanabe2013widely}.

In particular, the regularization prior of FBART is specified similarly to that of BART:
\begin{equation}\label{eq:fbart:prior}
	\pi\Big(\{\mT_k,\sM_k\}_{k=1}^K,\sigma^2\Big) =\pi(\sigma^2)\prod_{k=1}^K\pi( \sM_k\mid\mT_k)\pi(\mT_k).
\end{equation}
For the prior distributions of $\sM_k$'s and $\sigma^2$, we use conjugate priors
\begin{equation}\label{eq:fbart:prior:mu_and_sigma}
    \sigma^2 \sim \nu\lambda/\chi^2_\nu, \quad\pi\big(\sM_k \mid\mT_k \big)  = \prod_{\ell=1}^{L_k}\sN(\vmu_{k\ell};\vmu_\mu,\mV_\mu),
\end{equation}
where $\chi^2_\nu$ stands for the Chi-square distribution with degrees of freedom $\nu$, and hyperparameters include $\vmu_\mu\in\R^J$, the covariance matrix $\mV_\mu\in \R^{J\times J}$, $\lambda\in \R^{+}$, and $\nu\in\mathbb{N}^+$. For the prior distributions of $\mT_k$'s, we employ the same tree prior described in \cref{subsec:reviewBART} except for the splitting probability $p_{\text{split}}(d)$. Unlike the specification in \cite{chipman1998bayesian,chipman2010bart}, the splitting probability for constructing $\pi(\mT_k)$ takes the following form
\begin{equation}\label{eq:splitprob}
    p_{\text{split}}(d) = a\gamma^{d},
\end{equation}
where $a\in(0,1]$ and $\gamma\in(0,1)$ are hyperparameters. This modification is motivated by \cite{rovckova2019theory} to ensure that $\pi(\mT_k)$ exhibits certain tail behaviors.

We present the following Lemma~\ref{lemma:fbartupdate} as the cornerstone for the subsequent posterior sampling algorithms. It basically shows that
both the full conditional distributions of $\{\sM_k\}$ and the marginal (conditional) likelihood over $\{\sM_k\}$ have closed forms. For a regression map $\Xi:[0,1]^p\to\mathcal{F}$, we write $\Xi(t;\vx) := \Xi(\vx)(t)$  and $\Xi(\vt_i;\vx_i)\equiv\big(\Xi(t_{i,1};\vx_i),\ldots,\Xi(t_{i,m_i};\vx_i)\big)\trans$ for $i=1,\ldots, n$.
\begin{lemma}\label{lemma:fbartupdate}
     Consider the function-on-scalar regression problem in \eqref{eq:fos} with regression map $\Xi_{\bbT,\bbM}$ and the FBART prior specified by \eqref{eq:fbart:prior}\textendash\eqref{eq:splitprob}. Let $\vphi(\vt_i)\in\R^{m_i\times J}$ denote the matrix of $\vphi$ evaluated at $\vt_i$, whose $j$th column is $(\phi_{j}(t_{i1}),\ldots,\phi_{j}(t_{im_i}))\trans$ for $j\in[J]$. For each $k\in[K]$, let $\mT_{(k)} = \{\mT_{k'}\}_{k'\neq k}$, $\sM_{(k)}= \{\sM_{k'}\}_{k'\neq k}$, and define the partial residuals
     $$\vr_i =  \vy_i - \sum_{k'=1,k'\neq k}^K\Xi_{\mT_{k'},\sM_{k'}}(\vt_i;\vx_i)\quad (i=1,\ldots,n).$$
     Then, it holds that: 
    \begin{itemize}
        \item[(i)] The full conditional distribution of the node parameters $\sM_k = \{\vmu_{k\ell}\}_{\ell=1}^{L_k}$ follows the normal distribution given below:
        \begin{equation*}
        \Pi_n\big(\sM_k\mid \vy_1,\ldots,\vy_n,\mT_k,\mT_{(k)},\sM_{(k)},\sigma^2\big)= \prod_{\ell=1}^{L_k} \sN(\vmu_{k\ell};\vmu_{\text{post}}^{k\ell}, \mV_{\text{post}}^{k\ell}),
    \end{equation*}
        
        where
\begin{equation}\label{eq:posterior:mu}
    \mV_{\text{post}}^{k\ell} =\Big[\mV_\mu^{-1} + \frac{1}{\sigma^2}\sum_{i:\vx_i\in D_k^{\ell}}\vphi\trans(\vt_i)\vphi(\vt_i)\Big]^{-1},\ \ \ \  \vmu_{\text{post}}^{k\ell} = \mV_{\text{post}}^{k\ell}\Big[\frac{1}{\sigma^2}\sum_{i:\vx_i\in D_k^{\ell}}\vphi\trans(\vt_i)\vr_i + \mV_\mu^{-1}\vmu_\mu\Big].
\end{equation}

    \item[(ii)] Given other parameters, the marginal likelihood over $\sM_k$ is
    $$p(\vy_1,\ldots,\vy_n\mid \mT_k,\mT_{(k)},\sM_{(k)},\sigma^2)= \prod_{\ell=1}^{L_k}p(\{\vr_i\}_{i:\vx_i\in D_k^{\ell}}\mid \sigma^2),$$ where $p(\{\vr_i\}_{i:\vx_i\in D_k^{\ell}}\mid \sigma^2)$ equals
\begin{equation}\label{eq:posterior:marginal}
\frac{(2\pi\sigma^2)^{-\frac{N_{k\ell}}{2}}|\mV_\mu|^{-1/2}}{|\mV_{\text{post}}^{k\ell}|^{-1/2}}\exp\Big[\frac{1}{2}(\vmu_{\text{post}}^{k\ell})\trans(\mV_{\text{post}}^{k\ell})^{-1}\vmu_{\text{post}}^{k\ell}
    -\frac{1}{2\sigma^2}\sum_{i:\vx_i\in D_k^{\ell}}\vr_i\trans\vr_i -\frac{1}{2}\vmu_\mu\trans\mV_\mu^{-1}\vmu_\mu 
    \Big],
\end{equation}
    and $N_{k\ell} = \sum_{i:\vx_i\in D_k^{\ell}} m_i$ is the number of observations in the $\ell$th subregion induced by $\mT_k$.
    \end{itemize}
\end{lemma}

To conduct posterior inference for FBART through
Markov chain Monte Carlo (MCMC), we propose a Bayesian backfitting algorithm by tailoring the existing implementations of BART. 
The conjugate Gibbs sampling is used for updating $\sigma^2$ and $\{\sM_k\}_{k=1}^K$, while the Metropolis–Hastings (MH) updates are employed for updating $\{\mT_k\}_{k=1}^K$. Specifically, the proposal distribution $q(\mT,\mT^*)$ includes four moves: \textit{Grow, Prune, Change} and \textit{Prior}, following the R packages \code{bartMachine} \citep{kapelner2016bartmachine} and \code{SoftBART} \citep{linero2018bayesian}. The proposed MCMC procedure is summarized in \cref{alg:mcmc:fbart}. {The primary computational bottleneck arises from sampling and marginalizing the leaf parameters. For a leaf corresponding to the $\ell$th subregion induced by $\mT_k$, the associated  computational complexity is generally $O\left(N_{k\ell}J^2+J^3\right)$.} Additional details on implementation and hyperparameter specifications are given in Section~
S.1 of the Supplementary Materials \citep{Cao2026functional}. 

\begin{algorithm}[t!]
    \SetKwInput{KwData}{Input}
    \caption{Bayesian backfitting MCMC algorithm for FBART}\label{alg:mcmc:fbart}
    \fontsize{9}{9} \selectfont
    \KwData{
        Data $\big\{(\vx_i,\{Y_i(t_{ij})\}_{j=1}^{m_i})\big\}_{i=1}^n$; B-splines $\{\vphi_j\}_{j=1}^J$; Hyperparameters $(K,\vmu_\mu,\mV_\mu, \nu,\lambda, a,\gamma)$; Number of iterations $\text{MC}_{\text{iter}}$.
    }
    \KwResult{Posterior samples.}
    \For{$i_{\text{iter}}\in[\text{MC}_{\text{iter}}]$}{
        \For{$k\in[K]$}{

            Calculate the partial residuals $\vr_i =  \vy_i - \sum_{k'=1,k'\neq k}^K\Xi_{\mT_{k'},\sM_{k'}}(\vt_i;\vx_i)$ for $i\in[n]$.
            
            \textbf{1. Update $\mT_k$}: 
            
            (i). Sample a new $\mT^*_k$ from the proposal distribution $q(\mT_k, \mT^*_k)$.
            
            (ii). Accept the new sample and update $\mT_k= \mT^*_k$ with probability
            \begin{equation}\label{eq:MHratio}
                \alpha\left(\mT_k, \mT^*_k\right)=\min \left\{\frac{q\left(\mT_k^*, \mT_k\right)}{q\left(\mT_k, \mT_k^*\right)} \frac{p\big(\vy_1,\ldots,\vy_n\mid \mT_k^*,\mT_{(k)},\sM_{(k)},\sigma^2\big) \pi\left(\mT_k^*\right)}{p\big(\vy_1,\ldots,\vy_n\mid \mT_k,\mT_{(k)},\sM_{(k)},\sigma^2\big) \pi\left(\mT_k\right)}, \quad 1\right\},
            \end{equation}
            where  $p\big(\vy_1,\ldots,\vy_n\mid \mT_k^*,\mT_{(k)},\sM_{(k)},\sigma^2\big)$ and $p\big(\vy_1,\ldots,\vy_n\mid \mT_k,\mT_{(k)},\sM_{(k)},\sigma^2\big)$ are calculated according to Equation~\eqref{eq:posterior:marginal}.
            
            \textbf{2. Update $\sM_k$}: For each $\ell\in[L_k]$,
            \begin{equation}\label{eq:mcmc:mu}
                \vmu_{k\ell}\sim\sN(\vmu_{\text{post}}^{k\ell}, \mV_{\text{post}}^{k\ell}),
            \end{equation}
            where $\vmu_{\text{post}}^{k\ell}$ and $\mV_{\text{post}}^{k\ell}$ are calculated according to Equation~\eqref{eq:posterior:mu}.
        }
        \textbf{3. Update $\sigma^2$}:
        \begin{equation*}\label{eq:mcmc:sigma2}
            \sigma^2 \sim\mathrm{InvGamma}\Big(\frac{\nu+N_n}{2},\quad\frac{\lambda\nu+\sum_{i=1}^n\|\vy_i - \Xi_{\bbT,\bbM}(\vt_i;\vx_i)\|_2^2}{2}\Big),
        \end{equation*}
        where $\mathrm{InvGamma}(a,b)$ stands for an inverse gamma distribution with density $p(x) \propto x^{-a-1}\exp(-b/x) $.
    }
\end{algorithm}

\section{Shape-Constrained FBART (S-FBART)}\label{subsec:shapeFBART}
In this section, we extend our proposed FBART to its shape-constrained variant that incorporates prior knowledge of functional responses. By leveraging the properties of B-splines, we can manipulate the spline coefficient vector to control the shape of their linear combination \citep[e.g.,][]{abraham2015bayesian,pya2015shape,wang2021shape}.  Here, we consider the commonly used shape constraints of response curves, including positivity, monotonicity, and convexity. The following Lemma~\ref{lemma:bspline:shape} shows how to impose these shape constraints by imposing linear constraints on the spline coefficients.
\begin{lemma}\label{lemma:bspline:shape}
    Let $\{\phi_{j}\}_{j=1}^J$ denote the B-spline basis functions of order $q\geq 1$, with knots $\xi_1=\xi_2=\ldots=\xi_q<\xi_{q+1}<\ldots<\xi_{J}<\xi_{J+1} = \xi_{J+2} =\ldots= \xi_{J+q}$. Given a basis coefficient vector $\vmu\in\R^J$ such that $\mD\vmu\geq \mathbf{0}$ for some matrix $\mD\in\R^{J'\times J}$ with $J'\leq J$, we have
    \begin{itemize}
        \item[(i)] $\vphi\trans\vmu$ is positive (non-negative) if $\mD = \mI_J$;
        \item[(ii)] $\vphi\trans\vmu$ is increasing if the $j$th row of $\mD\in\R^{(J-1)\times J}$ is
        \begin{equation*}
            (0,\ldots,0,-1,1,0,\ldots,0),
        \end{equation*}
        where the indices of nonzero entries are $j$ and $j+1$, for $j\in[J-1]$;
        \item[(iii)] $\vphi\trans\vmu$ is convex if the $j$th row of $\mD\in\R^{(J-2)\times J}$ is
        \begin{equation*}
			\Big(0,\ldots,0,(\xi_{j+q} - \xi_{j+1})^{-1},  -(\xi_{j+q} - \xi_{j+1})^{-1} -(\xi_{j+q+1} - \xi_{j+2})^{-1} ,   (\xi_{j+q+1} - \xi_{j+2})^{-1} ,0,\ldots,0\Big),
		\end{equation*}
        where the indices of nonzero entries are $j$, $j+1$ and $j+2$, for $j\in[J-2]$.
    \end{itemize}    
\end{lemma}

We refer to the matrix $\mD$ in Lemma~\ref{lemma:bspline:shape} as the \textit{constraint matrix} for a given shape constraint. By combining different constraint matrices, we can impose more complex shape constraints on the fitted function $\vphi\trans\vmu$, such as \textit{both} monotonicity and convexity. See Section~
S.1.4 in the Supplementary Materials \citep{Cao2026functional} for more details.

Next, we discuss the posterior inference of the S-FBART model. Based on the above discussion, we extend the prior distribution of FBART given in \cref{subsec:prior_and_mcmc} to the one ensuring a required shape constraint. This extension is based on a constrained version of normal distributions. Given a constraint matrix $\mD$ corresponding to a certain shape constraint in Lemma~\ref{lemma:bspline:shape}, we say a random vector $\vmu \in \R^J$ follows a \textit{shape-constrained normal distribution} $\sN^{\mD}(\vmu_\mu, \mV_\mu)$, if its density has the following form:
\begin{equation*}
    p(\vmu) = \frac{1}{C_{\mD}(\vmu_\mu, \mV_\mu)\sqrt{(2\pi)^J|\mV_\mu|}} \exp\Big[-\frac{1}{2}(\vmu-\vmu_\mu)\trans\mV_\mu^{-1}(\vmu-\vmu_\mu)\Big]\mathbb{I}_{\{\vmu:\mD\vmu\geq \mathbf{0}\}}(\vmu),
\end{equation*}
where $C_{\mD}(\vmu_\mu, \mV_\mu)$ is a normalizing constant depending on the mean vector $\vmu_\mu$ and covariance matrix $\mV_\mu$. The shape-constrained normal distribution is closely related to the truncated normal distribution. In particular, let $\widebar\mD\in\R^{J\times J}$ denote an invertible matrix whose first $J'$ rows are $\mD$. By writing $\veta = \widebar\mD\vmu$, we have
\begin{equation}\label{eq:normalconstrain:transform}
    \vmu\sim\sN^{\mD}(\vmu_\mu, \mV_\mu) \Longleftrightarrow \veta  \sim \sN^{+}_{1:J'}(\widebar\mD\vmu_\mu,\widebar\mD\mV_\mu\widebar\mD\trans),
\end{equation}
where $\sN^{+}_{1:J'}$ denotes the truncated normal distribution with positivity constraints on the first $J'$ entries.

Given a constraint matrix $\mD$, the prior distribution of S-FBART, $\pi^\mD(\cdot)$, 
is defined by replacing the priors of $\{\sM_k\}$ in FBART with shape-constrained normal distributions:
\begin{equation}\label{eq:shapefbart:prior}
	\pi^\mD\Big(\{\mT_k,\sM_k\}_{k=1}^K,\sigma^2\Big) =\pi(\sigma^2)\prod_{k=1}^K \Big[\prod_{\ell=1}^{L_k}\sN^{\mD}(\vmu_{k\ell}; \vmu_\mu,\mV_\mu)\Big]\pi(\mT_k).
\end{equation}
\begin{corollary}
    In S-FBART, the induced prior and posterior distributions of $\Xi(\vx)$ satisfy the specified shape constraint for all $\vx\in[0,1]^p$.
\end{corollary}

Similar to Lemma~\ref{lemma:fbartupdate}, we present some basic results for S-FBART in Lemma~\ref{lemma:sfbartupdate}. To sample from the posterior, we use \cref{alg:mcmc:fbart} with two modifications: i) To update $\mT_k$, the marginal likelihood in Equation~\eqref{eq:MHratio} is calculated according to Equation~\eqref{eq:posterior:marginal:constrain} instead of Equation~\eqref{eq:posterior:marginal}; and ii) to update $\sM_k$ in Equation~\eqref{eq:mcmc:mu}, we sample from the shape-constrained normal distribution $\vmu_{k\ell}\sim \sN^{\mD}(\vmu_{\text{post}}^{k\ell}, \mV_{\text{post}}^{k\ell})$ in Equation~\eqref{eq:mcmc:mu:constrain}, for $\ell\in[L_k]$.

\begin{lemma}\label{lemma:sfbartupdate}
     Given a certain shape constraint in Lemma~\ref{lemma:bspline:shape} and the associated constraint matrix $\mD$, consider the function-on-scalar regression problem in \eqref{eq:fos} with regression map $\Xi_{\bbT,\bbM}$ and the S-FBART prior specified in Equation~\eqref{eq:shapefbart:prior}. For each $k\in[K]$, we have:
    \begin{itemize}
        \item[(i)] The full conditional distribution of the node parameters $\sM_k = \{\vmu_{k\ell}\}_{\ell=1}^{L_k}$ is 
        \begin{equation}\label{eq:mcmc:mu:constrain}
        \Pi_n\big(\sM_k\mid \vy_1,\ldots,\vy_n,\mT_k,\mT_{(k)},\sM_{(k)},\sigma^2\big)= \prod_{\ell=1}^{L_k} \sN^{\mD}(\vmu_{k\ell};\vmu_{\text{post}}^{k\ell}, \mV_{\text{post}}^{k\ell}),
    \end{equation}
    where $\vmu_{\text{post}}^{k\ell}$ and $\mV_{\text{post}}^{k\ell}$ are given in Equation~\eqref{eq:posterior:mu};
    \item[(ii)] The marginal likelihood $p^{\mD}(\vy_1,\ldots,\vy_n\mid \mT_k,\mT_{(k)},\sM_{(k)},\sigma^2)$ for S-FBART is
\begin{equation}\label{eq:posterior:marginal:constrain}
p(\vy_1,\ldots,\vy_n\mid \mT_k,\mT_{(k)},\sM_{(k)},\sigma^2)\times\prod_{\ell=1}^{L_k} \frac{C_{\mD}(\vmu_{\text{post}}^{k\ell}, \mV_{\text{post}}^{k\ell})}{C_{\mD}(\vmu_\mu, \mV_\mu)},
\end{equation}
    where $p(\vy_1,\ldots,\vy_n\mid \mT_k,\mT_{(k)},\sM_{(k)},\sigma^2)$ is given in Equation~\eqref{eq:posterior:marginal}.
    \end{itemize}
\end{lemma}

\begin{remark}
As shown in Equation~\eqref{eq:normalconstrain:transform}, the implementation of S-FBART involves sampling from truncated normal distributions and evaluating multivariate normal probabilities $C_{\mD}(\cdot,\cdot)$. Sampling from a truncated normal distribution can be achieved through methods such as rejection sampling or Gibbs sampling \citep[e.g.,][]{kotecha1999gibbs}, while normal integrals can be numerically computed using Monte Carlo algorithms \citep[e.g.,][]{genz2009computation}. Recently, \cite{botev2017normal} introduced a minimax tilting method that offers exact sampling and accurate integral calculation for truncated normal distributions. For S-FBART, we observe that a moderately large dimension (e.g., $J=10$) is sufficient to achieve the desired estimation and prediction accuracies in both simulation and real-data analyses, thereby avoiding the computational burden associated with high-dimensional truncated normal distributions.
\end{remark}

\section{Posterior Concentration Results}\label{sec:convergence}

In this section, we investigate the theoretical properties of FBART and S-FBART. Specifically, we establish consistency and derive posterior contraction rates for the proposed methods. Throughout this section, the covariate dimension $p$ is considered to be fixed for simplicity, and extension to high-dimensional regression is possible by introducing a sparsity-inducing prior \citep[e.g.,][]{linero2018bayesianHD}; we also fix the error variance $\sigma^2$ at $1$, noting that it can be generalized to an unknown $\sigma^2$ \citep{ghosal2017fundamentals}. For any two sequences $A_n$ and $B_n$, we write $A_n\lesssim B_n$ if $A_n\leq cB_n$ for some constant $c>0$ independent of $n$, $A_n\gtrsim B_n$ if $B_n\lesssim A_n$, and $A_n\asymp B_n$ if $A_n\lesssim B_n$ and $B_n\lesssim A_n$.

We consider observations $\big\{(\vx_i,\{Y_i(t_{ij})\}_{j=1}^{m_i})\big\}_{i=1}^n$ generated according to the FOSR model in \eqref{eq:fos}, and impose proper smoothness restriction on the true regression map $\Xi_0$. Recall $\Xi_0$ is a mapping from the Euclidean space $\mathbb{R}^p$ to the function space $\mathcal{F}$. Smoothness property is required for the mapping $\Xi_0$ itself as well as its functional output. In particular, $\Xi_0$ is assumed to belong to the following space:
\begin{equation*}
   \mathcal{HC}^{\alpha,\beta} := \Big\{\Xi: [0,1]^p\to C^{\alpha}[0,1];\ \  \sup_{\vx\neq \vx'}\frac{\|\Xi(\vx)-\Xi(\vx')\|_{C^{\alpha}}}{\|\vx-\vx'\|_2^{\beta}}<\infty\Big\},
\end{equation*}
where $\alpha>0$, $\beta\in(0,1]$, and $\|\cdot\|_{C^{\alpha}}$ denotes the H\"older norm of order $\alpha$. The parameter $\alpha$ regulates the smoothness of the functional output, and $\beta$ controls the smoothness of the mapping with respect to its vector input. 

The convergence results will be derived with respect to the following empirical metric:
\begin{equation*}
    d^2_n(\Xi,\Xi') := \frac{1}{N_n}\sum_{i=1}^n \|\Xi(\vt_i;\vx_i)-\Xi'(\vt_i;\vx_i)\|_2^2,
\end{equation*}
where $\Xi$ and $\Xi'$ are two regression maps, $N_n = \sum_{i=1}^n m_i$, and $m_i$ is the number of observed points for subject $i$. We allow each $m_i$ to (implicitly) depend on $n$, and assume that there exists a positive constant $\xi<\infty$ such that $(\max_{i=1}^n m_i)/(\min_{i=1}^n m_i)\leq \xi$ for all $n$. 

Let $\sG = \bigl\{\Xi : \Xi = \sum_{k=1}^K \Xi_{\mT_k,\sM_k}\bigr\}$ denote the space of all functional additive regression tree maps with a fixed number of trees $K$. For simplicity, we assume that the B-spline basis functions are of a fixed order $q\geq\alpha$, with equally spaced knots. We place a prior on $\sG$ by assigning prior distributions to the model parameters, namely the binary decision trees $\{\mT_k\}$, the node parameters $\{\sM_k\}$, and the basis dimension $J$:
\begin{equation}\label{eq:FBART:prior:theory}
    \pi_n\Big(\{\mT_k,\sM_k\}_{k=1}^K, J\Big) = \pi_n(J) \prod_{k=1}^K \pi_n(\mT_k \mid J)
    \prod_{\ell=1}^{L_k} \sN\bigl(\vmu_{k\ell}; \mathbf{0},\, \mI_J/K\bigr).
\end{equation}
Here, $\pi_n(\mT\mid J)$ follows the tree prior in \cite{chipman2010bart} with a splitting probability
\begin{equation}\label{eq:prior:thm:split}
    p_{\mathrm{split}}(d) \asymp \gamma^{J\log(N_n) + d},\quad \forall d\in\mathbb{N},
\end{equation}
where $\gamma \in(0,\tfrac{1}{2})$. This specification is motivated by \cite{rovckova2019theory} and has been further tailored for the FOSR problem by incorporating its dependence on $J$ and $N_n$. Moreover, we assume that the prior $\pi_n(J)$ satisfies
\begin{equation}\label{eq:prior:thm:J}
    \log \pi_n(J)\asymp -J(\log J)^{r},\quad \forall J\in\mathbb{N},
\end{equation}
 where $r\geq0$ is a constant. This condition holds for several well-known distributions, such as the geometric and Poisson distributions.


Our proof focuses on the space partition induced by k-d trees. A binary decision tree $\mT$ is called a k-d tree \citep{rovckova2020posterior} if it satisfies the following properties: 1) All the terminal nodes have the same depth; 2) the splitting variable cycles over $[p]$, and the internal nodes at the same depth share the same splitting variable; 3) the splitting value at each node is the median observed value in the node along the splitting variable. Based on the definition of the k-d tree, after $s$ rounds of splitting cycles, the resulting k-d tree has $L = 2^{sp}$ terminal nodes and each terminal node contains at least $\lfloor n/L\rfloor$ observations. The induced partition $\sD = \{D^1,\ldots,D^L\}$ by a k-d tree is referred to as a k-d tree partition.

To proceed, we assume that the design points $\{\vx_i\}_{i=1}^n$ are ``regular'' as in Condition~\ref{condition:design} below. Intuitively, Condition~\ref{condition:design} requires the design points $\{
\vx_i\}_{i=1}^n$ be approximately uniform in the predictor space (e.g., a regular grid on $[0, 1]^p$ satisfies this condition). 

\begin{condition}\label{condition:design}
    There exists a constant $M>0$ such that for any $s\geq 1$, the k-d tree partition $\sD = \{D^1,\ldots,D^L\}$ with $L=2^{sp}$ satisfies
    \begin{equation*}
        \max_{1\leq \ell \leq L}\mathrm{diam}(D^{\ell}) \leq M\sum_{\ell=1}^L \frac{n_{\ell}}{n} \mathrm{diam}(D^{\ell}),
    \end{equation*}
    where $\mathrm{diam}(D^{\ell}) = \max\limits_{\vx_i,\vx_{i'}\in {D}^{\ell}}\|\vx_i-\vx_{i'}\|_2$ and $n_{\ell} = \sum\limits_{i=1}^n\mathbb{I}_{D^{\ell}}(\vx_i)$.
\end{condition}

The following Lemma~\ref{lemma:approximation} gives the error bound of the k-d tree map for approximating the true functional regression map. 

\begin{lemma}\label{lemma:approximation}
    Assume $\Xi_0\in\mathcal{HC}^{\alpha,\beta}$ for some $\alpha>0$ and $\beta\in(0,1]$, and $\{\vx_i\}_{i=1}^n$ satisfies Condition~\ref{condition:design}. Let $\vphi = (\phi_1,\ldots,\phi_J)\trans$ be a set of B-spline basis functions of order $q\geq\alpha$ with equally spaced knots.
    Then, for any k-d tree $\mT$ with $L$ terminal nodes, there exists a set of node parameters $\widehat{\sM}$ such that the tree-structured step map $\widehat\Xi = \Xi_{\mT,\widehat{\sM}}$ satisfies
    \begin{equation*}
	d_n(\widehat\Xi,\Xi_0)  \lesssim J^{-\alpha} + L^{-\beta/p}.
	\end{equation*}
\end{lemma}
The proof is provided in Section~S.4.1 of the Supplementary Materials \citep{Cao2026functional}. 

Our posterior convergence results rely on three conditions to hold \citep[e.g., see][]{ghosal2007convergence,rovckova2020posterior}, which are presented in detail in Section~S.4 
of the Supplementary Materials \citep{Cao2026functional}. The primary challenges for verifying these conditions are to derive the prior concentration rate at $\Xi_0$, and to properly construct subsets $\sG_n\subseteq\sG$ that can well approximate $\sG$ with relatively low complexity. The following main theorem establishes the posterior consistency of our proposed FBART estimator. 

\begin{theorem}\label{thm:convergence:forest}
    Assume $\Xi_0\in\mathcal{HC}^{\alpha,\beta}$ for some $\alpha>0$ and $\beta\in(0,1]$, and Condition~\ref{condition:design} is satisfied. Let the space $\sG$ be endowed with the FBART prior specified in Equations~\eqref{eq:FBART:prior:theory}--\eqref{eq:prior:thm:J}. Then with $\varepsilon_n = N_n^{-\alpha\beta/\{\alpha(2\beta+p) + \beta\}}\log^{1/2} N_n$, we have
\begin{equation*}\label{eq:thm1:rate}
        \Pi_n\Big(\Xi \in \sG:d_n(\Xi,\Xi_0)> C_n \varepsilon_n \mid Y_1(\vt_1),\ldots, Y_n(\vt_n)\Big) \longrightarrow 0
    \end{equation*}
    for any $C_n\to \infty$ in $\Prob^n_{\Xi_0}$-probability, as $n\to\infty$.
\end{theorem}
The proof is provided in Section~S.4.4 of the Supplementary Materials \citep{Cao2026functional}. 

\begin{remark}
The above theoretical result is rate-adaptive in the sense that the FBART prior does not rely on the unknown smoothness parameters $(\alpha,\beta)$ of the regression map $\Xi_0$. In particular, the proof of \cref{thm:convergence:forest} reveals that the ``best'' dimension $J$, which balances the squared bias and variance, satisfies $J \asymp N_n^{\beta/\{\alpha(2\beta + p) + \beta\}}$. Moreover, we observe that larger values of $\alpha$ or $\beta$ lead to a faster contraction rate $\varepsilon_n$, aligning with intuition.
\end{remark}

For S-FBART proposed in \cref{subsec:shapeFBART}, we have similar convergence results. We first investigate how the linear constraint in Lemma~\ref{lemma:bspline:shape} affects the approximation power of B-spline functions. 
\begin{lemma}\label{lemma:bspline:approximation:constrain}
We define a function $Y(t)\in C^{\alpha}[0,1]$ to be $\kappa$-strictly shape-constrained for some $\kappa>0$, if $Y(t)$ satisfies one of the following conditions:
\begin{itemize}
    \item[(i)] $Y(t)$ is strictly positive, i.e., $Y(t)\geq \kappa$ for all $t\in[0,1]$;
    \item[(ii)] $Y(t)$ is strictly increasing with $\alpha> 1$, i.e., $\frac{\mathrm{d}Y(t)}{\mathrm{d}t}\geq \kappa$  for all $t\in[0,1]$;
    \item[(iii)] $Y(t)$ is strictly convex with $\alpha> 2$, i.e., $\frac{\mathrm{d}^2}{\mathrm{d}t^2}Y(t)\geq \kappa$  for all $t\in[0,1]$.
\end{itemize}
Let $\vphi = (\phi_1,\ldots,\phi_J)\trans$ be a set of B-spline basis functions of order $q\geq\alpha$ with equally spaced knots, and $\mD$ be the associated constraint matrix for $Y(t)$. Then, for large enough $J$, we have 
\begin{equation*}
\inf_{\vmu\in\R^J:\mD\vmu\geq \mathbf{0}}\|\vphi\trans\vmu - Y\|_{\infty} \lesssim J^{-\alpha}.
\end{equation*}
\end{lemma}
The proof is provided in Section~S.4.5 of the Supplementary Materials \citep{Cao2026functional}. 

Lemma~\ref{lemma:bspline:approximation:constrain} shows that the approximation error of the constrained B-splines representation is of the same order as that of the unconstrained one, as long as the target function is strictly shape-constrained. When the shape constraints are not strict, the corresponding best approximation error can be sub-optimal \citep{de1974splines}. The following theorem establishes the consistency of the S-FBART estimator for shape-constrained functional responses. 

\begin{theorem}\label{thm:forest:constrain}
    Under the same conditions and settings as described in \cref{thm:convergence:forest}, suppose in addition that there exists $\kappa>0$ such that $\Xi_0(\vx)$ is $\kappa$-strictly shape-constrained for all $\vx\in[0,1]^p$ with constraint matrix $\mD$. Let the space $\sG$ be endowed with the following S-FBART prior:
    \begin{equation*}
        \pi_n^{\mD}\Big(\{\mT_k,\sM_k\}_{k=1}^K, J\Big) = \pi_n(J) \prod_{k=1}^K \pi_n(\mT_k \mid J)
    \prod_{\ell=1}^{L_k} \sN^{\mD}\bigl(\vmu_{k\ell}; \mathbf{0},\, \mI_J/K\bigr).
\end{equation*}
    Then, the contraction result~\eqref{eq:thm1:rate} in \cref{thm:convergence:forest} holds for the S-FBART posterior with the same rate $\varepsilon_n = N_n^{-\alpha\beta/\{\alpha(2\beta+p) + \beta\}}\log^{1/2} N_n$.
\end{theorem}
The proof is provided in Section~S.4.5 of the Supplementary Materials \citep{Cao2026functional}. 

\section{Numerical Studies}\label{sec:simu}

\subsection{Simulation setup}\label{subsec:simu:setup}
{We evaluate the performance of the proposed FBART and S-FBART methods through simulation studies. We consider the model in \eqref{eq:fos} with $p=2$ covariates and functional responses defined on $[0,1]$. We independently generate $n=400$ covariate vectors from a uniform distribution over the covariate space $[0,1]^p$. Each response curve, $\{Y_i(t_{ij})\}_{j\in[m_i]}$, is observed at $m_i=m=20$ sampling points. We use $J=10$ cubic B-spline basis functions with equally spaced knots to approximate the true response function and consider two noise levels, $\sigma\in\{0.1,1\}$. Unless stated otherwise, all simulation settings in this section follow this baseline specification. 

In this section, three sampling designs for $t_{ij}$ are considered to provide a comprehensive assessment of the methods' performance: i) a regular-grid design, with $t_{ij}=\frac{j}{m_i+1}$ for $j\in[m_i]$, representing the common setting in which all functional responses are observed on a shared regular grid; ii) a uniform random design, with $t_{ij}\overset{\mathrm{i.i.d.}}{\sim}\mathrm{Uniform}([0,1])$; and iii) an irregular random design, with $t_{ij}\overset{\mathrm{i.i.d.}}{\sim}\mathrm{Uniform}([0,0.25]\cup[0.75,1])$, mimicking functional data subject to a partially observed domain.} 

For the true regression map, we consider the following {four} cases:

Case 1: A piece-wise constant map 
\begin{equation*}
    \Xi_1(t;\vx) = \big\{1+\bbI_{[0,0.5]}(\vx(1))\big\}\tan\Big(4\pi\big\{t + t^{2\bbI_{[0,0.5]}(\vx(2))} - 1\big\}/9\Big);
\end{equation*}

Case 2: A map involving both smooth and non-smooth parts:
\begin{equation*}
    \Xi_2(t;\vx) =\frac{4\vx(1)+1}{\vx(2) + 2}t + 2\big\{\vx(2)\bbI_{[0,0.5]}(\vx(1)) + \vx(2)\big\}\tan\Big(4\pi\big[t + t^{4\vx(2)} - 1\big]/9\Big);
\end{equation*}

Case 3: A linear map:
\begin{equation*}
    \Xi_3(t; \vx) = \vx(1) + \vx(2) + \big\{1+2\vx(1) + 4\vx(2)\big\}\Phi^{-1}(t),
\end{equation*}
where $\Phi^{-1}(\cdot)$ is the quantile function of the standard normal distribution. For Cases 1-3, the true regression map is monotonically increasing in $t$ for any $\vx\in[0,1]^2$;

{Case 4: A nonlinear map with convex functional  responses:
\begin{equation*}
    \Xi_4(t;\vx) =\frac{4\vx(1)+1}{\vx(2) + 2}t + 10\big\{\vx(2)\bbI_{[0,0.5]}(\vx(1)) + \vx(2)\big\}\Big((1.5\bbI_{[0.5,1]}(\vx(2))+0.5)t^2 - \vx(1)t\Big).
\end{equation*}}

We compare the proposed FBART and S-FBART with several state-of-the-art competing methods, including the classical BART \citep{chipman2010bart}, the monotone BART \citep[mBART,][]{chipman2022mbart}, the BART with targeted smoothing \citep[tsBART,][]{starling2020bart} and its monotone version \citep[tsBART-m,][]{starling2019monotone}, the Bayesian FOSR method \citep[BFOSR,][]{kowal2020bayesian}, and the local linear regression method with functional responses \citep[LLR, e.g.,][]{petersen2019frechet,fan2022conditional}. For BART and mBART, we treat $Y_i(t_{ij})$ as the response value with the covariate vector $(t_{ij},\vx_i\trans)\trans$ of dimension $(p+1)$. A monotonically increasing constraint in $t$ is imposed for mBART and tsBART-m {in Cases 1--3}. For the proposed S-FBART method, we use the constraint matrix $\mD$ corresponding to the monotonically increasing constraint {in Cases 1--3 and the convexity constraint in Case 4, as} defined in Lemma~\ref{lemma:bspline:shape}. 
For LLR, we use the normal kernel function, with the bandwidth chosen to minimize the in-sample five-fold cross-validation root mean squared error. {All benchmark methods are configured using the default or recommended hyperparameter settings provided by their respective reference software implementations.}

The prediction performance of different methods is quantified by three metrics. The first metric is the root mean squared prediction errors (RMSPE) defined as $\textsc{rmspe} = \sqrt{\frac{1}{N_n^*}\sum_{i=1}^{n^*}\|\hat\Xi(\vt^*_i;\vx^*_i)-\Xi_0(\vt^*_i;\vx^*_i)\|_2^2}$, {where $N_n^* = \sum_{i=1}^{n^*}m_i^*$, $\vx^*_i\in[0,1]^2$ and $\vt^*_i\in [0,1]^{m_i^*}$ denote the covariate vector and the vector of sampling points, respectively, for the $i$-th observation in a test sample of size $n^*$}; for Bayesian approaches, we use the posterior mean for point estimation. In addition, we calculate the pointwise posterior $95\%$ credible interval for uncertainty quantification. {The accuracy of the credible interval estimates for the Bayesian approaches (all methods except LLR)} is evaluated via the mean negatively oriented interval score \citep[MIS,][]{gneiting2007strictly}, defined as $\textsc{mis} = \frac{1}{N_n^*}\sum_{i=1}^{n^*}\sum_{j=1}^{m_i^*}\Big[\hat U_{ij}-\hat L_{ij} + \frac{2}{5\%}\inf_{\eta\in [\hat L_{ij},\hat U_{ij}]} |\Xi_0(t^*_{ij};\vx^*_i)- \eta|\Big]$, where $\hat U_{ij}$ and $\hat L_{ij}$ are the $97.5\%$-quantile and $2.5\%$-quantile of the posterior samples of $\Xi(t^*_{ij};\vx^*_i)$, respectively. Last, we use the mean continuous ranked probability score \citep[MCRPS,][]{gneiting2007strictly} to evaluate the performance of probabilistic prediction. For each simulation setup, these three metrics are evaluated {on a test sample of size $n^*=400$, with each response observed at $m^*_i=m$ sampling points. The test sampling points are set to $t_{ij}^*=j/(m_i^*+1)$ for $j\in[m_i^*]$, under the regular-grid design and are independently drawn from $\mathrm{Uniform}([0,1])$ under both random designs.} For all three metrics, a lower value indicates a better performance.

\subsection{Simulation results}\label{Sec_Sim_results}

{First, we independently generate $20$ replicates for each simulation setting described in \cref{subsec:simu:setup} under the regular-grid design.} The average RMSPE, MIS, and MCRPS values for all methods are shown in Table~\ref{table:simu:main}; the results of mBART and tsBART-m are not reported for Case 4 since they are not applicable to convex functional responses. For each metric, the first two best results are shown in \textbf{bold}. For {Cases 1, 2, and 4}, where the true relationship between the functional response and the covariate exhibits non-linearity and lack of smoothness, our proposed FBART and S-FBART methods consistently outperform other methods in terms of prediction performance and uncertainty quantification. This is not very surprising because FBART and S-FBART account for both the functional nature of the responses and the non-linearity of the true regression map.  {The competing BART-based methods (i.e., BART, mBART, tsBART, and tsBART-m) generally outperform BFOSR and LLR, which assume a linear or locally linear relationship between the response and the covariates. However, these BART-based competitors do not adequately capture the functional structure of the responses, resulting in inferior predictive performance relative to FBART and S-FBART.} 
For Case 3, where a linear regression map is assumed, both BFOSR and LLR outperform other methods. This outcome aligns with our expectations, since these two methods are specifically designed for (locally) linear regression maps. However, it is noteworthy that FBART and S-FBART still manage to achieve MCRPSs comparable to those of linear models. 

Across all simulation scenarios, FBART and S-FBART consistently outperform the four BART-based competing methods in terms of all three metrics. The superior performance of  FBART and S-FBART can be attributed to the spline modeling of functional data, which effectively captures the inherent functional nature of responses.


\begin{table}[ht!]
\caption{Comparison results of different methods in terms of three metrics.}\label{table:simu:main}
{
\centering
\resizebox{\textwidth}{!}{%
\begin{tabular}{c|c|c|cccccccc}
    \hline\hline
      $\sigma$& Case&  Metric & FBART & S-FBART & BART & mBART  & tsBART & tsBART-m & BFOSR & LLR \\ 
      \hline
\multirow{12}{*}{$1$} & \multirow{3}{*}{case 1} 
& RMSPE  & $\tb{0.177}$&$\tb{0.175}$&$0.480$&$0.257$&$0.341$&$0.428$&$0.893$&$0.524$\\
& & MIS   & $\tb{0.623}$&$\tb{0.840}$&$5.974$&$1.597$&$2.133$&$3.374$&$5.958$&-- \\ 
& & MCRPS   & $\tb{0.574}$&$\tb{0.574}$&$0.624$&$0.582$&$0.597$&$0.614$&$0.771$&$0.637$\\
\cline{2-11}
& \multirow{3}{*}{case 2} 
& RMSPE  & $\tb{0.236}$&$\tb{0.215}$&$0.839$&$0.335$&$0.356$&$0.502$&$0.806$&$0.367$\\
& & MIS   & $\tb{1.100}$&$\tb{0.994}$&$13.931$&$3.695$&$2.020$&$3.752$&$10.798$&-- \\ 
& & MCRPS   & $\tb{0.580}$&$\tb{0.578}$&$0.731$&$0.595$&$0.600$&$0.629$&$0.888$&$0.601$\\

\cline{2-11}
& \multirow{3}{*}{case 3} 
& RMSPE  &$0.182$&$0.166$&$0.621$&$0.235$&$0.311$&$0.346$&$\tb{0.050}$&$\tb{0.084}$\\
& & MIS   & $\tb{0.891}$&$\tb{0.918}$&$10.479$&$2.118$&$1.908$&$2.300$&$1.400$&--\\ 
& & MCRPS   & $0.575$&$0.573$&$0.667$&$0.581$&$0.592$&$0.598$&$\tb{0.567}$&$\tb{0.568}$\\ 
\cline{2-11}
& \multirow{3}{*}{case 4}
& RMSPE & $\tb{0.474}$ & $\tb{0.501}$ & $0.868$ & -- & $0.789$ & -- & $2.538$ & $1.049$ \\
& & MIS & $\tb{1.671}$ & $\tb{2.031}$ & $10.241$ & -- & $7.630$ & -- & $26.771$ & --\\
& & MCRPS & $\tb{0.601}$ & $\tb{0.610}$ & $0.714$ & -- & $0.709$ & -- & $1.970$ & $0.762$ \\
\hline
\multirow{12}{*}{$0.1$} & \multirow{3}{*}{case 1} 
& RMSPE  &$\tb{0.165}$&$\tb{0.172}$&$0.496$&$0.257$&$0.212$&$0.232$&$0.890$&$0.350$\\
& & MIS   & $\tb{0.474}$&$\tb{0.504}$&$8.856$&$2.212$&$1.637$&$1.785$&$7.165$&--\\
 
& & MCRPS   & $\tb{0.072}$&$\tb{0.073}$&$0.242$&$0.098$&$0.111$&$0.117$&$0.512$&$0.154$\\

\cline{2-11}
& \multirow{3}{*}{case 2} 
& RMSPE  &$\tb{0.139}$&$\tb{0.142}$&$0.806$&$0.210$&$0.263$&$0.257$&$0.803$&$0.207$\\
& & MIS   & $\tb{0.720}$&$\tb{0.706}$&$17.359$&$3.407$&$3.051$&$2.940$&$3.986$&--\\
& & MCRPS   & $\tb{0.092}$&$\tb{0.092}$&$0.451$&$0.109$&$0.145$&$0.143$&$0.401$&$0.094$\\

\cline{2-11}
& \multirow{3}{*}{case 3} 
& RMSPE  &$0.049$&$0.043$&$0.583$&$0.094$&$0.098$&$0.095$&$\tb{0.007}$&$\tb{0.008}$\\
& & MIS   & $\tb{0.282}$&$0.289$&$14.015$&$2.080$&$0.786$&$0.748$&$\tb{0.132}$&--\\
& & MCRPS   & $0.064$&$0.063$&$0.365$&$0.077$&$0.076$&$0.075$&$\tb{0.057}$&$\tb{0.057}$\\
\cline{2-11}
& \multirow{3}{*}{case 4}
& RMSPE & $\tb{0.450}$ & $\tb{0.500}$ & $0.899$ & -- & $0.655$ & -- & $2.528$ & $0.913$ \\
& & MIS & $\tb{1.637}$ & $\tb{1.567}$ & $13.809$ & -- & $7.827$ & -- & $11.783$ & -- \\
& & MCRPS & $\tb{0.177}$ & $\tb{0.182}$ & $0.383$ & -- & $0.311$ & -- & $1.269$ & $0.320$ \\
\hline\hline
\end{tabular}
}
}

\end{table}

Next, we compare FBART and BART with their respective shape-constrained counterparts. We observe significant improvements when transitioning from BART to mBART across all scenarios. On the other hand, S-FBART yields comparable performance to FBART, {with neither method consistently outperforming the other}; a similar pattern is also observed for tsBART and tsBART-m. A possible explanation is that incorporating the shape information is particularly beneficial for BART, since it does not make use of the functional structure of the responses. 


{To further investigate the effect of shape constraints, we compare FBART and S-FBART under two random sampling designs with $\sigma=1$. We first consider the uniform random design, in which both the training sampling points $t_{ij}$ and test sampling points $t^*_{ij}$ are independently drawn from $\mathrm{Uniform}([0,1])$. Figure~\ref{fig:ratio:random} presents side-by-side boxplots of the FBART-to-S-FBART ratios for RMSPE and MCRPS in Cases 2 and 4, with $m\in\{5,10,20\}$. A ratio greater than one indicates better performance of S-FBART. Across all values of $m$, S-FBART slightly outperforms FBART on both metrics in Case~2 but underperforms FBART in Case~4, although the differences between the two methods are generally modest. This observation is consistent with the corresponding results in Table~\ref{table:simu:main}. Thus, when the training and test curves follow the same sampling distribution, neither method uniformly dominates the other. In this setting, the main benefits of imposing shape constraints are ensuring that the estimated functions satisfy the specified shape restrictions and improving interpretability, rather than consistently improving predictive performance.

While the uniform random design does not reveal a consistent predictive advantage from imposing shape constraints, the benefit becomes much clearer under the irregular random design, where the training and test sampling distributions differ. We now fix $m=10$ and consider the irregular random design, in which the training points satisfy $t_{ij}\overset{\mathrm{i.i.d.}}{\sim}\mathrm{Uniform}([0,0.25]\cup[0.75,1])$, whereas the test points $t^*_{ij}$ are independently sampled from $\mathrm{Uniform}([0,1])$. This design leaves the interior interval $[0.25,0.75]$ unobserved during training. We evaluate the predictive performance over the full test domain and separately over the training region, $[0,0.25]\cup[0.75,1]$, and the extrapolation region, $[0.25,0.75]$. The corresponding metric ratios are reported in Figure~\ref{fig:ratio:random:gap}. Under this distribution shift, S-FBART shows a substantial overall advantage over FBART, driven primarily by markedly better performance in the extrapolation region, while the two methods remain comparable over the observed training region. This suggests that enforcing shape constraints stabilizes function estimates over unobserved intervals, leading to improved extrapolation accuracy.

\begin{figure}[ht!]
    \centering
    \includegraphics[width=0.8\linewidth]{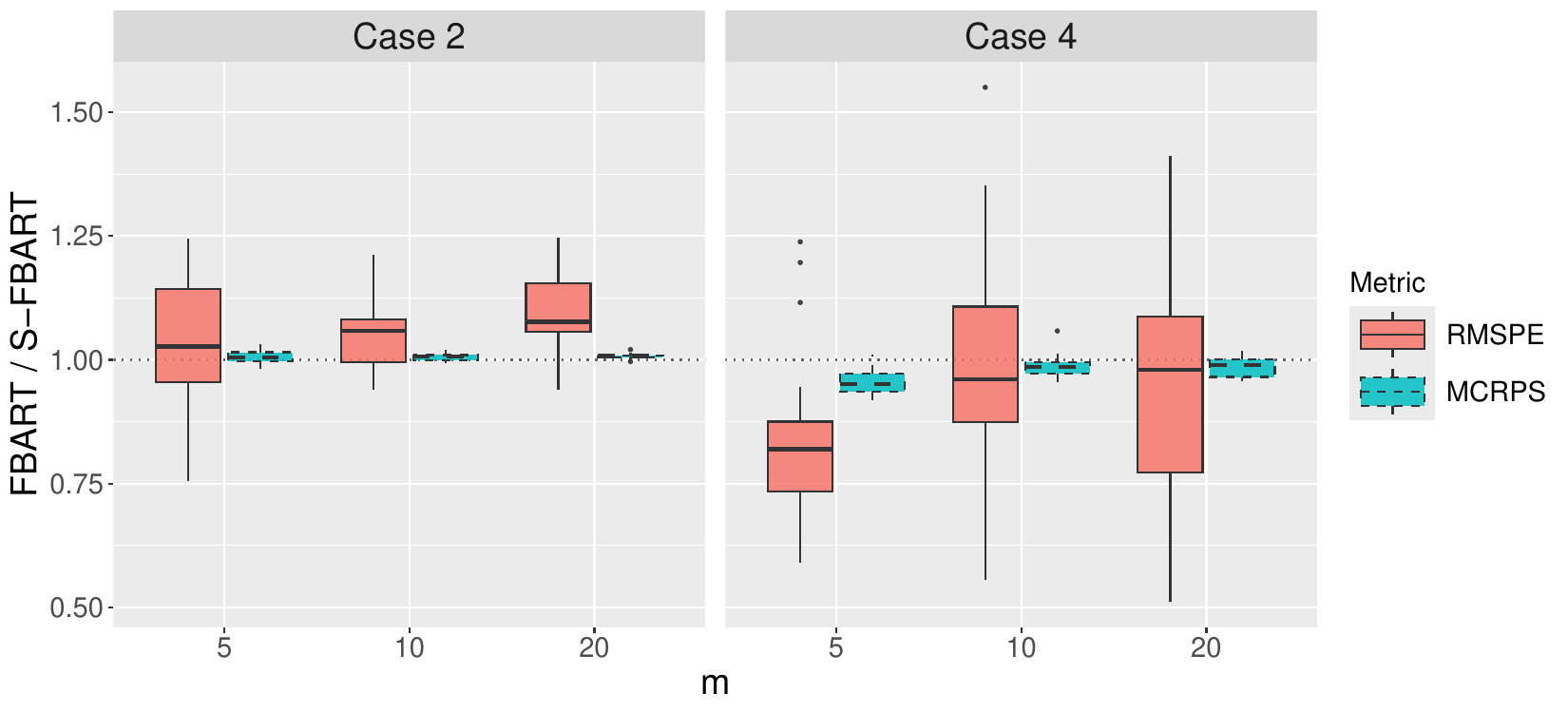}
    \caption{Ratios of the RMSPE and MCRPS values for FBART relative to S-FBART under the uniform random sampling design.}
    \label{fig:ratio:random}
\end{figure}

\begin{figure}[ht!]
    \centering
    \includegraphics[width=0.8\linewidth]{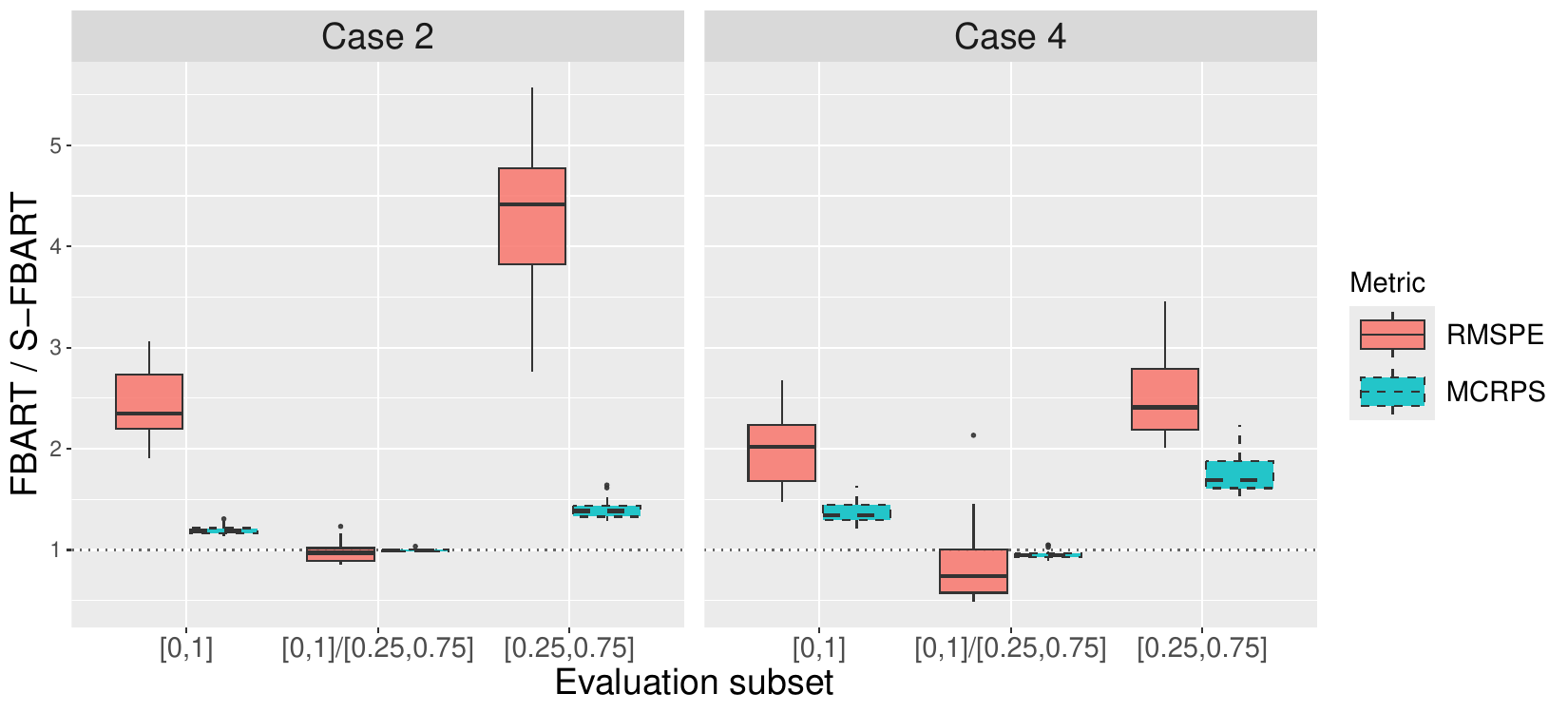}
    \caption{Ratios of the RMSPE and MCRPS values for FBART relative to S-FBART under the irregular random sampling design, with training sampling points $t_{ij}$ sampled uniformly from $[0,0.25]\cup[0.75,1]$ and test sampling points $t^{*}_{ij}$ sampled uniformly from $[0,1]$. Results are reported for the full domain, the training region, and the extrapolation region $[0.25,0.75]$.}
    \label{fig:ratio:random:gap}
\end{figure}
}

\begin{figure}[h!]
    \centering
    \includegraphics[width=\linewidth]{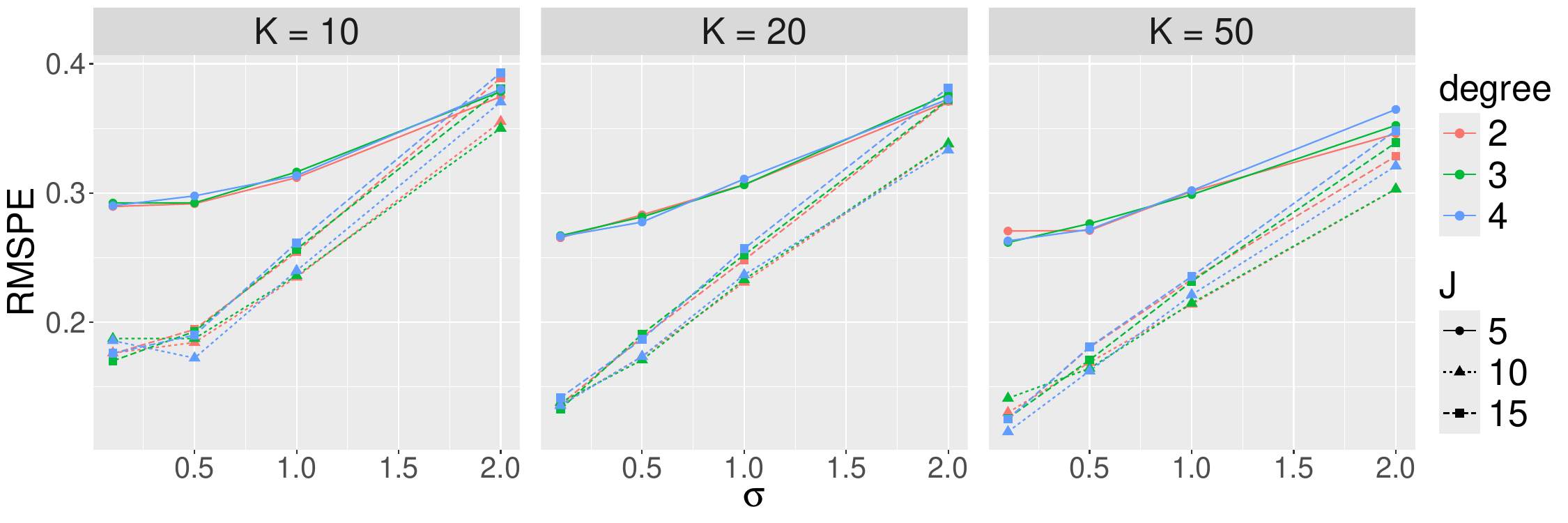}
    \caption{The performance of FBART under different tuning parameter specifications.}
    \label{fig:simu:FBART}
\end{figure}

Finally, we examine the performance of FBART under a variety of noise levels and tuning parameter selections {under the regular-grid design}. Specifically, we consider the true regression map in Case 2, with noise level $\sigma\in\{0.1,0.5,1,2\}$,  the {order} of B-spline basis $q\in \{2,3,4\}$, the number of trees $K\in\{10,20,50\}$, and the number of basis functions $J\in\{5,10,15\}$. Figure~\ref{fig:simu:FBART} shows the average RMSPEs of FBART based on $10$ simulation runs. As expected, the resulting RMSPE scales approximately linearly with $\sigma$. In addition, we observe that choosing a larger $K$ can improve the prediction accuracy, and empirically using $K=20$ trees is sufficient to deliver prediction results comparable to those of using larger numbers of trees. With respect to the dimension $J$ of the basis functions, we find that a moderately large dimension ($J=10$) is adequate. In Section~S.2.3 
of the Supplementary Materials \citep{Cao2026functional}, it is shown that the MCRPS results of FBART are also quite robust to different specifications of tuning parameters. 

{ More extensive simulation results, including analyses under varying sample sizes, basis specifications, and covariate dimensions, as well as computational time comparisons and detailed implementation descriptions, are provided in Section~S.2 of the Supplementary Materials.}

\section{Real Data Illustrations}\label{sec:RD}

The proposed FBART and S-FBART methods are applied to two real datasets, \code{Battery} and \code{Wage}, each exhibiting distinct shape constraints on their response curves. For comparison, we also present prediction results from BART, mBART, and tsBART, as described in \cref{sec:simu}. The predictive performance of all methods is assessed on test datasets using root mean squared prediction error (RMSPE), mean absolute prediction error (MAPE), and mean continuous ranked probability score (MCRPS). In real data applications, where the true response values are unknown, RMSPE and MAPE are computed by comparing the predictions to the observed outcomes.

For FBART and S-FBART, the optimal number of trees, $K \in \{1,10,20\}$, and the basis dimension, $J \in \{5,10,15\}$, are selected by minimizing the widely applicable information criterion \citep[WAIC,][]{watanabe2013widely}. For S-FBART, we impose a monotonicity constraint on the \code{Battery} dataset and a concavity constraint on the \code{Wage} dataset. For BART, mBART, and tsBART, the number of trees is similarly determined from the set {$\{1,10,20,50, 200\}$} according to WAIC. {All other specifications are set to their respective default values.} We exclude mBART from the analysis of the \code{Wage} dataset, since the response curves do not exhibit a monotonic behaviour. In addition, tsBART is not applied to the \code{Battery} dataset due to the substantial computational demands of its Gaussian process component. Finally, BFOSR and LLR in \cref{sec:simu} are not considered here because their available implementations cannot accommodate curves observed at irregular sampling locations. Implementation details and additional results are given in Section~S.3 
of the Supplementary Materials \citep{Cao2026functional}.

\subsection{Data description}\label{subsec:data}

Given the significant concern of energy challenges in modern society, accurate prediction of battery performance is crucial for battery production and optimization. The first dataset, \code{Battery}, contains capacity values of $124$ lithium-ion batteries cycled under fast-charging conditions \citep{severson2019data}. Our target is to predict the battery's capacity fade curve, where battery capacity is treated as a function of the number of charge-discharge cycles. Following \cite{severson2019data}, the prediction starts from the $101$th cycle onward, with $p = 9$ features constructed from the early-cycle data (the data in the cycles from $1$ to $100$) as the covariates. We randomly select $93$ curves for training, with the remaining $31$ curves for testing. A logit transformation is performed on the capacity values to make the Gaussian noise assumption more applicable. For the capacity fade curves, it is reasonable to assume that they are monotonically decreasing in cycle numbers (see Figure S.10
).

Economists have long been interested in studying the impact of various variables on individual incomes \citep[e.g.,][]{card1999causal,rubinstein2006post}. The second dataset, \code{Wage}, contains weekly wages of full-time working males in the United States in 1987 \citep[see the data object \code{ex2019} in the R package \code{Sleuth2};][]{ramsey2002statistical}. Here we explore the relationship between the wage curve (wages versus work experience) and workers' features, including years of education, whether the person is black,  whether the workplace is in a city, and the region of the workplace (i.e., $p=4$). We randomly select $n = 15,000$ samples for training and the remaining $10,437$ samples for testing. Previous studies \citep[e.g.,][]{hannah2013multivariate,chernina2023wages} have suggested a concave relationship between wages and the years of work experience.

\subsection{Results}

The prediction results of the two real datasets are summarized in Table~\ref{table:realdata}, with the best results highlighted in \textbf{bold} font. We also include a ``Decreasing'' column for \code{Battery} and a ``Concave'' column for \code{Wage} to indicate whether a method accounts for the shape information of response curves. We observe that for both datasets, the proposed FBART and S-FBART models generally outperform the competing methods in terms of prediction accuracy and probabilistic prediction. 
This demonstrates the benefits of accounting for responses' functional nature. 
Notably, S-FBART offers flexibility in incorporating various shape constraints, whereas mBART is limited to modeling monotone curves.

\begin{table}[!ht]
\caption{Prediction results of different methods on the \code{Battery} and \code{Wage} datasets.}
\label{table:realdata}
\centering
\resizebox{\textwidth}{!}{%
\begin{tabular}{l|cccc|cccc}
\hline
& \multicolumn{4}{c|}{\code{Battery}} & \multicolumn{4}{c}{\code{Wage}} \\
\cline{2-9}
Method & RMSPE & MAPE & MCRPS & Decreasing
       & RMSPE & MAPE & MCRPS & Concave \\
\hline
FBART
& $0.050$ & $0.025$ & $0.156$ & \xmark
& $371.886$ & $236.532$ & $180.741$ & \xmark \\

S-FBART
& $\tbi{0.037}$ & $\tbi{0.019}$ & $\tbi{0.128}$ & \checkmark
& $\tbi{371.499}$ & $\tbi{236.328}$ & $\tbi{180.592}$ & \checkmark \\

BART
& $0.049$ & $0.029$ & $0.333$ & \xmark
& $374.684$ & $239.401$ & $181.094$ & \xmark \\

mBART
& $0.067$ & $0.037$ & $0.443$ & \checkmark
& -- & -- & -- & -- \\

tsBART
& -- & -- & -- & --
& $372.493$ & $237.068$ & $180.935$ & \xmark \\
\hline\hline
\end{tabular}%
}
\end{table}

{When comparing S-FBART with FBART, we find that S-FBART achieves comparable predictive performance on the \code{Wage} dataset but substantially improves both prediction accuracy and uncertainty quantification on the \code{Battery} dataset, as indicated by its lower RMSPE, MAPE, and MCRPS values. One possible explanation is that, in the \code{Wage} dataset, the training and test samples cover similar ranges of years of experience. In contrast, the battery capacity-fade curves are observed over markedly different cycle ranges due to early termination of battery cycling, leading to a greater shift in the functional domain between the training and test data. In this setting, incorporating shape information may improve extrapolation to sparsely observed or previously unobserved regions of the sampling domain.}

\begin{figure}[!ht]
    \centering
    \includegraphics[width=\linewidth]{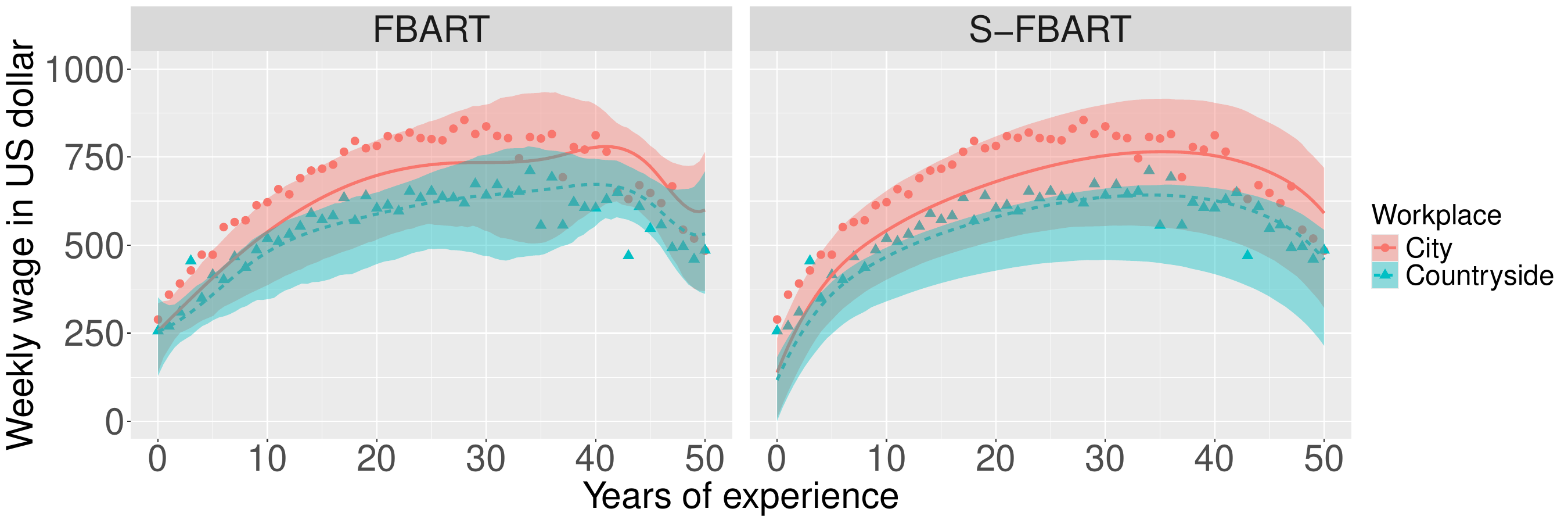}
    \caption{Posterior wage curves for synthetic individuals working in a city and outside a city using FBART and S-FBART. Points denote the average observed wage at each year of experience within the corresponding subpopulation, representing the empirical wage profile of each synthetic individual. Curves show the posterior mean wage functions, and shaded areas indicate the corresponding pointwise $95\%$ credible regions.}
    \label{fig:RD:wage}
\end{figure}

Finally, we use the \code{Wage} dataset to illustrate the effect of imposing shape constraints. We consider two synthetic ``representative'' individuals, with one working in a city (``City'') and the other working outside a city (``Countryside''); the covariates of these two synthetic individuals are set to the covariate values averaged over their respective subpopulations. Figure~\ref{fig:RD:wage} shows the estimation results of FBART and S-FBART with $K=J=10$ for these two synthetic individuals. We observe that both methods produce posterior distributions that align well with the observed data. Notably, S-FBART produces strictly concave posterior samples, while FBART yields curves of irregular shape and greater uncertainty, failing to satisfy the expected concave-shape constraint.

\section{Discussion}\label{sec:discussion}

We have proposed a flexible Bayesian approach for the function-on-scalar regression (FOSR) problem with potential shape constraints on the response curves. To the best of our knowledge, FBART is the first Bayesian tree-based method for the FOSR problem, and S-FBART is the first theoretically backed-up Bayesian nonparametric approach for the shape-constrained FOSR. The proposed methods are especially preferred when dealing with a highly nonlinear regression map defined on a complex covariate space. 

The results presented above show that the proposed Functional BART (FBART) and its shape-constrained variant (S-FBART) provide a flexible and effective solution for function-on-scalar regression. By leveraging spline-based functional representations and a Bayesian tree ensemble, our approach is able to model highly nonlinear and heterogeneous relationships that are often encountered in real-world functional data. Incorporating shape constraints further enhances estimation and prediction when prior knowledge on response curves is available. {The benefits are twofold. First, compared with FBART, S-FBART achieves similar estimation accuracy while producing substantially narrower credible intervals (see Section~S.2 of the Supplementary Materials). Second, S-FBART demonstrates better extrapolation performance than FBART and may therefore be particularly useful when the distribution of sampling points differs between the training and test data.} 
{Both simulation studies and real-data applications demonstrate that FBART and S-FBART provide accurate predictions and reliable uncertainty quantification at a reasonable computational cost, particularly for moderate-to-large datasets with complex nonlinear functional regression relationships.} Notably, our methods maintain strong theoretical guarantees, with posterior contraction rates that adapt to the unknown smoothness of the underlying regression map. Together, these findings highlight the practical utility and theoretical robustness of the proposed framework for modern functional data analysis.

{The proposed function-on-scalar regression framework in Equation~\eqref{eq:fos} focuses on estimating the conditional expectation $\mathbb{E}(Y \mid \mathbf{x})$ from observations drawn from the joint distribution of $(Y, \mathbf{x})$. From a functional data analysis perspective, $Y \in \mathcal{F}$ is treated as a random function, where $\mathcal{F}$ may incorporate structural properties such as smoothness, monotonicity, or convexity. Unlike functional models that explicitly parameterize within-curve correlation, our formulation assumes independent Gaussian white-noise residuals as a working simplification, concentrating the functional structure on the conditional mean $\Xi(t; \mathbf{x})$. Similar modeling strategies are widely employed in the recent functional data analysis literature \citep[e.g.,][]{kowal2020bayesian, kundu2025decomposition}. This independence assumption can be readily relaxed within the FBART framework. For instance, $\epsilon_i(t)$ in Equation~\eqref{eq:fos} may be modeled as a Gaussian process with a parametric covariance kernel, with the associated covariance parameters updated within the MCMC algorithm using Metropolis--Hastings steps. Furthermore, the Gaussian white-noise specification admits a generalized Bayesian interpretation: conditional on $\sigma^2$, the posterior distribution coincides with a generalized posterior based on an empirical squared $L_2$ loss. From this point of view, the independent Gaussian error model acts as a quasi-likelihood that targets the conditional mean function, rather than a generative model that completely specifies the full within-curve dependence structure \citep{knoblauch2022optimization}.

In this work, the proposed shape-constrained priors are defined through linear inequalities on the B-spline leaf parameters, yielding convex polyhedral parameter spaces that naturally accommodate standard global constraints such as positivity, monotonicity, and convexity. Consequently, S-FBART is particularly well suited to functional shapes characterized by linear derivative restrictions. Localized shape restrictions over a subinterval may also be incorporated by constraining only the B-spline coefficients associated with that region, taking advantage of the local support of the basis functions. In contrast, more general non-linear or non-convex shape restrictions, such as requiring a specific number of peaks and valleys, cannot be directly accommodated by the current specification and would require new prior constructions and posterior sampling strategies.}

The proposed FBART and S-FBART methods integrate infinite-dimensional data with Bayesian tree-based models, opening several promising directions for future research. One natural extension is to consider function-on-function regression, which would require designing efficient domain partitioning models tailored to functional spaces. Another important direction is the extension to multivariate functional data or image data. This could be achieved by incorporating multivariate basis functions, such as tensor product B-splines or thin plate splines. In terms of shape constraints, it would be practically valuable to allow for different types of constraints across curves and to develop data-driven methods for inferring the appropriate constraint type. Furthermore, S-FBART has potential applications beyond traditional functional data, such as modeling quantile functions or probability distributions, which naturally exhibit shape constraints like monotonicity.

The theoretical results in this work can also be expanded in several directions. First, it would be insightful to derive minimax convergence rates over various classes of regression maps and to investigate whether FBART achieves these optimal rates. {Minimax theory has received relatively limited attention in function-on-scalar regression, and available results suggest that imposing shape constraints often does not alter the convergence rate relative to the unconstrained setting \citep[e.g.,][]{mammen1999smoothing,meyer2008inference,petersen2019frechet,kundu2025decomposition}. Developing rigorous posterior contraction theory for shape-constrained tree ensembles under specific settings, such as distribution shift, and deriving the corresponding minimax rates represent interesting directions for future theoretical work. It is also worth noting that, as the smoothness of the response functions increases with $\alpha\to\infty$, our contraction rate approaches $N_n^{-\beta/(2\beta+p)}$ up to logarithmic factors, closely resembling the classical minimax-optimal rate for nonparametric regression. Furthermore, although our main theoretical results assume $\beta\in(0,1]$, the contraction-rate theory could potentially be extended to $\beta>1$ by incorporating smooth gating functions, following the softBART construction \citep{linero2018bayesian}.} Another important direction is to establish Bernstein–von Mises results, which would provide insight into the asymptotic behavior of the posterior distribution and offer frequentist justification for the resulting Bayesian inference, particularly in the presence of shape priors.

\begin{acks}[Acknowledgments]
We thank the Editor, the Associate Editor, and two anonymous reviewers for their insightful and constructive comments that greatly helped us to improve the manuscript. Bohai Zhang and Shiyuan He contributed equally as
corresponding authors of this manuscript.
\end{acks}

\begin{funding}


B.~Zhang was partially supported by the Natural Science Foundation of Guangdong Province, China (2025A1515010213), the Zhuhai Basic and Applied Basic Research Foundation Project (2320004002570), Guangdong and Hong Kong Universities ``1+1+1" Joint Research Collaboration Scheme (2025A0505000010), the Guangdong Provincial Key Laboratory of IRADS (2022B1212010006) and the UIC start-up fund (UICR0700083-24). S.~He was partially supported by the National Natural Science Foundation of China (Grant Numbers 12571278).
\end{funding}

\begin{supplement}
\stitle{Technical material}
\sdescription{The file (Supp\_FBART.pdf, PDF file) provides technical details of the theoretical results and algorithm implementation. The file also contains supplementary results of the numerical studies.}
\end{supplement}
\begin{supplement}
\stitle{Code}
\sdescription{The archive file (SupplementaryCode\_FBART.zip) contains the R code implementing the proposed method with several illustrative examples. A README file is included to describe its contents.}
\end{supplement}

\bibliographystyle{chicago}
\bibliography{ref}

@article{he2023unified,
  title={A unified analysis of multi-task functional linear regression models with manifold constraint and composite quadratic penalty},
  author={He, Shiyuan and Ye, Hanxuan and He, Kejun},
  journal={Journal of Machine Learning Research},
  volume={24},
  number={291},
  pages={1--69},
  year={2023}
}

@book{deboor1978practical,
  title = {A Practical Guide to Splines},
  author = {Carl de Boor},
  series = {Applied Mathematical Sciences},
  publisher = {Springer New York, NY},
  copyright = {Springer-Verlag New York 1978},
  year = {1978},
  note = {Published: 29 November 2001},
  edition = {1},
  pages = {XVIII, 348}
}

@article{pya2015shape,
  title={Shape constrained additive models},
  author={Pya, Natalya and Wood, Simon N},
  journal={Statistics and computing},
  volume={25},
  pages={543--559},
  year={2015},
  publisher={Springer}
}

@article{lin2014bayesian,
  title={Bayesian monotone regression using Gaussian process projection},
  author={Lin, Lizhen and Dunson, David B},
  journal={Biometrika},
  volume={101},
  number={2},
  pages={303--317},
  year={2014},
  publisher={Oxford University Press}
}

@article{kowal2020bayesian,
  title={Bayesian function-on-scalars regression for high-dimensional data},
  author={Kowal, Daniel R and Bourgeois, Daniel C},
  journal={Journal of Computational and Graphical Statistics},
  volume={29},
  number={3},
  pages={629--638},
  year={2020},
  publisher={Taylor \& Francis}
}

@article{chipman2010bart,
author = {Hugh A. Chipman and Edward I. George and Robert E. McCulloch},
title = {{BART: Bayesian additive regression trees}},
volume = {4},
journal = {The Annals of Applied Statistics},
number = {1},
publisher = {Institute of Mathematical Statistics},
pages = {266 -- 298},
year = {2010},
doi = {10.1214/09-AOAS285},
}

@article{hill2020bayesian,
  title={Bayesian additive regression trees: {A} review and look forward},
  author={Hill, Jennifer and Linero, Antonio and Murray, Jared},
  journal={Annual Review of Statistics and Its Application},
  volume={7},
  pages={251--278},
  year={2020},
  publisher={Annual Reviews}
}

@article{denison1998bayesian,
	title={A bayesian cart algorithm},
	author={Denison, David GT and Mallick, Bani K and Smith, Adrian FM},
	journal={Biometrika},
	volume={85},
	number={2},
	pages={363--377},
	year={1998},
	publisher={Oxford University Press}
}

@article{li2023adaptive,
  title={Adaptive conditional distribution estimation with Bayesian decision tree ensembles},
  author={Li, Yinpu and Linero, Antonio R and Murray, Jared},
  journal={Journal of the American Statistical Association},
  volume={118},
  number={543},
  pages={2129--2142},
  year={2023},
  publisher={Taylor \& Francis}
}

@article{chipman2022mbart,
  title={mBART: multidimensional monotone BART},
  author={Chipman, Hugh A and George, Edward I and McCulloch, Robert E and Shively, Thomas S},
  journal={Bayesian Analysis},
  volume={17},
  number={2},
  pages={515--544},
  year={2022},
  publisher={International Society for Bayesian Analysis}
}

@article{rovckova2020posterior,
  title={Posterior concentration for Bayesian regression trees and forests},
  author={Ro{\v{c}}kov{\'a}, Veronika and Van der Pas, Stephanie},
  journal={The Annals of Statistics},
  volume={48},
  number={4},
  pages={2108--2131},
  year={2020},
  publisher={JSTOR}
}

@article{ghosal2007convergence,
  title={Convergence rates of posterior distributions for noniid observations},
  author={Ghosal, S and van der Vaart, AW},
  journal={Annals of Statistics},
  volume={35},
  number={1},
  pages={192--223},
  year={2007},
  publisher={Institute of Mathematical Statistics}
}

@article{ge2019random,
  title={Random tessellation forests},
  author={Ge, Shufei and Wang, Shijia and Teh, Yee Whye and Wang, Liangliang and Elliott, Lloyd},
  journal={Advances in Neural Information Processing Systems},
  volume={32},
  year={2019}
}

@article{chipman1998bayesian,
  title={Bayesian CART model search},
  author={Chipman, Hugh A and George, Edward I and McCulloch, Robert E},
  journal={Journal of the American Statistical Association},
  volume={93},
  number={443},
  pages={935--948},
  year={1998},
  publisher={Taylor \& Francis}
}

@article{luo2021bast,
  title={BAST: Bayesian additive regression spanning trees for complex constrained domain},
  author={Luo, Zhao Tang and Sang, Huiyan and Mallick, Bani},
  journal={Advances in Neural Information Processing Systems},
  volume={34},
  pages={90--102},
  year={2021}
}

@article{linero2018bayesian,
  title={Bayesian regression tree ensembles that adapt to smoothness and sparsity},
  author={Linero, Antonio R and Yang, Yun},
  journal={Journal of the Royal Statistical Society Series B: Statistical Methodology},
  volume={80},
  number={5},
  pages={1087--1110},
  year={2018},
  publisher={Oxford University Press}
}

@article{um2023bayesian,
  title={Bayesian additive regression trees for multivariate skewed responses},
  author={Um, Seungha and Linero, Antonio R and Sinha, Debajyoti and Bandyopadhyay, Dipankar},
  journal={Statistics in Medicine},
  volume={42},
  number={3},
  pages={246--263},
  year={2023},
  publisher={Wiley Online Library}
}

@article{morris2015functional,
  title={Functional regression},
  author={Morris, Jeffrey S},
  journal={Annual Review of Statistics and Its Application},
  volume={2},
  pages={321--359},
  year={2015},
  publisher={Annual Reviews}
}

@article{birke2007estimating,
  title={Estimating a convex function in nonparametric regression},
  author={Birke, Melanie and Dette, Holger},
  journal={Scandinavian Journal of Statistics},
  volume={34},
  number={2},
  pages={384--404},
  year={2007},
  publisher={Wiley Online Library}
}

@article{hannah2013multivariate,
  title={Multivariate convex regression with adaptive partitioning},
  author={Hannah, Lauren A and Dunson, David B},
  journal={The Journal of Machine Learning Research},
  volume={14},
  number={1},
  pages={3261--3294},
  year={2013},
  publisher={JMLR. org}
}

@article{ghosal2023shape,
  title={Shape-constrained estimation in functional regression with Bernstein polynomials},
  author={Ghosal, Rahul and Ghosh, Sujit and Urbanek, Jacek and Schrack, Jennifer A and Zipunnikov, Vadim},
  journal={Computational Statistics \& Data Analysis},
  volume={178},
  pages={107614},
  year={2023},
  publisher={Elsevier}
}

@book{ramsey2002statistical,
  title = {The Statistical Sleuth: A Course in Methods of Data Analysis},
  author = {Ramsey, Fred L. and Schafer, Daniel W.},
  year = {2002},
  publisher = {Duxbury/Thomson Learning},
  edition = {2nd}
}

@article{de1974splines,
  title={Splines with nonnegative {B}-spline coefficients},
  author={De Boor, Carl and Daniel, James W},
  journal={Mathematics of Computation},
  volume={28},
  number={126},
  pages={565--568},
  year={1974}
}

@article{kapelner2016bartmachine,
  title={bartMachine: Machine Learning with Bayesian Additive Regression Trees},
  author={Kapelner, Adam and Bleich, Justin},
  journal={Journal of Statistical Software},
  volume={70},
  pages={1--40},
  year={2016}
}

@article{abraham2015bayesian,
  title={Bayesian regression with {B-splines} under combinations of shape constraints and smoothness properties},
  author={Abraham, Christophe and Khadraoui, Khader},
  journal={Statistica Neerlandica},
  volume={69},
  number={2},
  pages={150--170},
  year={2015},
  publisher={Wiley Online Library}
}

@book{genz2009computation,
  title={Computation of multivariate normal and t probabilities},
  author={Genz, Alan and Bretz, Frank},
  volume={195},
  year={2009},
  publisher={Springer Science \& Business Media}
}

@inproceedings{kotecha1999gibbs,
  title={Gibbs sampling approach for generation of truncated multivariate gaussian random variables},
  author={Kotecha, Jayesh H and Djuric, Petar M},
  booktitle={1999 IEEE international conference on acoustics, speech, and signal processing. Proceedings. ICASSP99 (Cat. No. 99CH36258)},
  volume={3},
  pages={1757--1760},
  year={1999},
  organization={IEEE}
}

@article{botev2017normal,
  title={The normal law under linear restrictions: simulation and estimation via minimax tilting},
  author={Botev, Zdravko I},
  journal={Journal of the Royal Statistical Society Series B: Statistical Methodology},
  volume={79},
  number={1},
  pages={125--148},
  year={2017},
  publisher={Oxford University Press}
}

@inproceedings{rovckova2019theory,
  title={On theory for BART},
  author={Ro{\v{c}}kov{\'a}, Veronika and Saha, Enakshi},
  booktitle={The 22nd international conference on artificial intelligence and statistics},
  pages={2839--2848},
  year={2019},
  organization={PMLR}
}

@book{ghosal2017fundamentals,
  title={Fundamentals of Nonparametric Bayesian Inference},
  author={Ghosal, Subhashis and Van der Vaart, Aad},
  volume={44},
  year={2017},
  publisher={Cambridge University Press}
}

@article{greven2017general,
  title={A general framework for functional regression modelling},
  author={Greven, Sonja and Scheipl, Fabian},
  journal={Statistical Modelling},
  volume={17},
  number={1-2},
  pages={1--35},
  year={2017},
  publisher={Sage Publications Sage India: New Delhi, India}
}

@article{liu2021variable,
  title={Variable selection with ABC Bayesian forests},
  author={Liu, Yi and Ro{\v{c}}kov{\'a}, Veronika and Wang, Yuexi},
  journal={Journal of the Royal Statistical Society Series B: Statistical Methodology},
  volume={83},
  number={3},
  pages={453--481},
  year={2021},
  publisher={Oxford University Press}
}

@book{ramsay2005functional,
  title = {Functional Data Analysis},
  author = {Ramsay, J. O. and Silverman, B. W. },
  series = {Springer Series in Statistics},
  publisher = {Springer New York, NY},
  edition = {2},
  pages = {XIX, 429},
  year = {2005},
  note = {Published: 08 June 2005, Softcover Published: 10 November 2010, eBook Published: 28 June 2006}
}

@article{fan2022conditional,
  title={Conditional distribution regression for functional responses},
  author={Fan, Jianing and M{\"u}ller, Hans-Georg},
  journal={Scandinavian Journal of Statistics},
  volume={49},
  number={2},
  pages={502--524},
  year={2022},
  publisher={Wiley Online Library}
}

@article{castillo2021uncertainty,
  title={Uncertainty quantification for Bayesian CART},
  author={Castillo, Ismael and Ro{\v{c}}kov{\'a}, Veronika},
  journal={The Annals of Statistics},
  volume={49},
  number={6},
  pages={3482--3509},
  year={2021},
  publisher={Institute of Mathematical Statistics}
}

@article{chiou2004functional,
  title={Functional response models},
  author={Chiou, Jeng-Min and M{\"u}ller, Hans-Georg and Wang, Jane-Ling},
  journal={Statistica Sinica},
  volume={14},
  pages={675--693},
  year={2004},
  publisher={JSTOR}
}

@article{zhang2022high,
  title={High-dimensional spatial quantile function-on-scalar regression},
  author={Zhang, Zhengwu and Wang, Xiao and Kong, Linglong and Zhu, Hongtu},
  journal={Journal of the American Statistical Association},
  volume={117},
  number={539},
  pages={1563--1578},
  year={2022},
  publisher={Taylor \& Francis}
}

@article{yao2005functional,
  title={Functional linear regression analysis for longitudinal data},
  author={Yao, Fang and M{\"u}ller, Hans-Georg and Wang, Jane-Ling},
  journal={Annals of Statistics},
  volume={33},
  number={6},
  pages={2873--2903},
  year={2005}
}

@article{scheipl2015functional,
	title={Functional additive mixed models},
	author={Scheipl, Fabian and Staicu, Ana-Maria and Greven, Sonja},
	journal={Journal of Computational and Graphical Statistics},
	volume={24},
	number={2},
	pages={477--501},
	year={2015},
	publisher={Taylor \& Francis}
}

@article{wang2021shape,
  title={Shape-Restricted Regression Splines with R Package splines2},
  author={Wang, Wenjie and Yan, Jun},
  journal={Journal of Data Science},
  volume={19},
  number={3},
  pages={498--517},
  year={2021}
}

@article{wang2016functional,
	title={Functional data analysis},
	author={Wang, Jane-Ling and Chiou, Jeng-Min and M{\"u}ller, Hans-Georg},
	journal={Annual Review of Statistics and Its Application},
	volume={3},
	pages={257--295},
	year={2016},
	publisher={Annual Reviews}
}

@article{ramsay1991some,
  title={Some Tools for Functional Data Analysis},
  author={Ramsay, J.O. and Dalzell, C.J.},
  journal={Journal of the Royal Statistical Society: Series B (Methodological)},
  volume={53},
  number={3},
  pages={539--561},
  year={1991},
  publisher={Wiley Online Library}
}

@article{morris2006wavelet,
  title={Wavelet-based functional mixed models},
  author={Morris, Jeffrey S and Carroll, Raymond J},
  journal={Journal of the Royal Statistical Society Series B: Statistical Methodology},
  volume={68},
  number={2},
  pages={179--199},
  year={2006},
  publisher={Oxford University Press}
}

@article{rosen2009bayesian,
	title={A Bayesian regression model for multivariate functional data},
	author={Rosen, Ori and Thompson, Wesley K},
	journal={Computational statistics \& data analysis},
	volume={53},
	number={11},
	pages={3773--3786},
	year={2009},
	publisher={Elsevier}
}

@article{chen2016variable,
	title={Variable selection in function-on-scalar regression},
	author={Chen, Yakuan and Goldsmith, Jeff and Ogden, R Todd},
	journal={Stat},
	volume={5},
	number={1},
	pages={88--101},
	year={2016},
	publisher={Wiley Online Library}
}

@article{linero2018bayesianHD,
  title={Bayesian regression trees for high-dimensional prediction and variable selection},
  author={Linero, Antonio R},
  journal={Journal of the American Statistical Association},
  volume={113},
  number={522},
  pages={626--636},
  year={2018},
  publisher={Taylor \& Francis}
}

@article{petersen2019frechet,
  title={Fr{\'e}chet regression for random objects with Euclidean predictors},
  author={Petersen, Alexander and M{\"u}ller, Hans-Georg},
  journal={The Annals of Statistics},
  volume={47},
  number={2},
  pages={691--719},
  year={2019},
  publisher={JSTOR}
}

@article{gneiting2007strictly,
  title={Strictly proper scoring rules, prediction, and estimation},
  author={Gneiting, Tilmann and Raftery, Adrian E},
  journal={Journal of the American Statistical Association},
  volume={102},
  number={477},
  pages={359--378},
  year={2007},
  publisher={Taylor \& Francis}
}

@article{card1999causal,
  title={The causal effect of education on earnings},
  author={Card, David},
  journal={Handbook of labor economics},
  volume={3},
  pages={1801--1863},
  year={1999},
  publisher={Elsevier}
}

@article{rubinstein2006post,
  title={Post schooling wage growth: Investment, search and learning},
  author={Rubinstein, Yona and Weiss, Yoram},
  journal={Handbook of the Economics of Education},
  volume={1},
  pages={1--67},
  year={2006},
  publisher={Elsevier}
}

@article{chernina2023wages,
  title={Do wages grow with experience? Deciphering the Russian puzzle},
  author={Chernina, Eugenia and Gimpelson, Vladimir},
  journal={Journal of Comparative Economics},
  volume={51},
  number={2},
  pages={545--563},
  year={2023},
  publisher={Elsevier}
}

@article{tang2008pairwise,
  title={Pairwise curve synchronization for functional data},
  author={Tang, Rong and M{\"u}ller, Hans-Georg},
  journal={Biometrika},
  volume={95},
  pages={875--889},
  year={2008},
  publisher={Oxford University Press}
}

@article{severson2019data,
  title={Data-driven prediction of battery cycle life before capacity degradation},
  author={Severson, Kristen A and Attia, Peter M and Jin, Norman and Perkins, Nicholas and Jiang, Benben and Yang, Zi and Chen, Michael H and Aykol, Muratahan and Herring, Patrick K and Fraggedakis, Dimitrios and others},
  journal={Nature Energy},
  volume={4},
  number={5},
  pages={383--391},
  year={2019},
  publisher={Nature Publishing Group UK London}
}

@article{unser1993b,
  title={B-spline signal processing. II. Efficiency design and applications},
  author={Unser, Michael and Aldroubi, Akram and Eden, Murray},
  journal={IEEE Transactions on Signal Processing},
  volume={41},
  number={2},
  pages={834--848},
  year={1993},
  publisher={IEEE}
}

@article{horowitz2017nonparametric,
  title={Nonparametric estimation and inference under shape restrictions},
  author={Horowitz, Joel L and Lee, Sokbae},
  journal={Journal of Econometrics},
  volume={201},
  number={1},
  pages={108--126},
  year={2017},
  publisher={Elsevier}
}

@book{groeneboom2014nonparametric,
  title={Nonparametric estimation under shape constraints},
  author={Groeneboom, Piet and Jongbloed, Geurt},
  number={38},
  year={2014},
  publisher={Cambridge University Press}
}

@article{wang2025monotone,
  title={Monotone Cubic B-Splines with a Neural-Network Generator},
  author={Wang, Lijun and Fan, Xiaodan and Li, Huabai and Liu, Jun S},
  journal={Journal of Computational and Graphical Statistics},
  pages={1--15},
  year={2025},
  publisher={Taylor \& Francis}
}

@article{starling2020bart,
  title={BART WITH TARGETED SMOOTHING},
  author={Starling, Jennifer E and Murray, Jared S and Carvalho, Carlos M and Bukowski, Radek K and Scott, James G},
  journal={The Annals of Applied Statistics},
  volume={14},
  number={1},
  pages={28--50},
  year={2020},
  publisher={JSTOR}
}

@article{watanabe2013widely,
  title={A widely applicable {Bayesian} information criterion},
  author={Watanabe, Sumio},
  journal={The Journal of Machine Learning Research},
  volume={14},
  number={1},
  pages={867--897},
  year={2013},
  publisher={JMLR. org}
}

@article{starling2019monotone,
  title={Monotone function estimation in the presence of extreme data coarsening: Analysis of preeclampsia and birth weight in urban Uganda},
  author={Starling, Jennifer E and Aiken, Catherine E and Murray, Jared S and Nakimuli, Annettee and Scott, James G},
  journal={arXiv preprint arXiv:1912.06946},
  year={2019}
}

@article{kowal2019functional,
  title={Functional autoregression for sparsely sampled data},
  author={Kowal, Daniel R and Matteson, David S and Ruppert, David},
  journal={Journal of Business \& Economic Statistics},
  volume={37},
  number={1},
  pages={97--109},
  year={2019},
  publisher={Taylor \& Francis}
}

@article{zhu2023statistical,
  title={Statistical learning methods for neuroimaging data analysis with applications},
  author={Zhu, Hongtu and Li, Tengfei and Zhao, Bingxin},
  journal={Annual Review of Biomedical Data Science},
  volume={6},
  number={1},
  pages={73--104},
  year={2023},
  publisher={Annual Reviews}
}

@article{hays2012functional,
  title={Functional dynamic factor models with application to yield curve forecasting},
  author={Hays, Spencer and Shen, Haipeng and Huang, Jianhua Z},
  journal={The Annals of Applied Statistics},
  pages={870--894},
  year={2012},
  publisher={JSTOR}
}

@article{jann2016estimating,
  title={Estimating Lorenz and concentration curves},
  author={Jann, Ben},
  journal={The Stata Journal},
  volume={16},
  number={4},
  pages={837--866},
  year={2016},
  publisher={SAGE Publications Sage CA: Los Angeles, CA}
}

@misc{he2025gs,
  title={{GS-BART: Bayesian additive regression trees with graph-split decision rules}},
  author={He, Shuren and Sang, Huiyan and Zhou, Quan},
  note={arXiv preprint arXiv:2509.07166},
  year={2025}
}

@article{luo2024nonstationary,
  title={A nonstationary soft partitioned Gaussian process model via random spanning trees},
  author={Luo, Zhao Tang and Sang, Huiyan and Mallick, Bani},
  journal={Journal of the American Statistical Association},
  volume={119},
  number={547},
  pages={2105--2116},
  year={2024},
  publisher={Taylor \& Francis}
}

@article{diaz2025nonparametric,
  title={Nonparametric Density Estimation on Complex Domains using Manifold-Aware {Bayesian} Additive Tree Models},
  author={Diaz-Ray, Isaac and Sang, Huiyan and Hu, Guanyu and Lu, Ligang},
  journal={Computational Statistics \& Data Analysis},
  pages={108335},
  year={2025},
  publisher={Elsevier}
}

@article{Cao2026functional,
  title={Supplement to ``{Functional BART with Shape Priors: A Bayesian Tree Approach to Constrained Functional Regression}"},
  author={Cao, Jiahao and He, Shiyuan and Zhang, Bohai},
  year={2026}
}

@article{knoblauch2022optimization,
  title={An optimization-centric view on Bayes' rule: Reviewing and generalizing variational inference},
  author={Knoblauch, Jeremias and Jewson, Jack and Damoulas, Theodoros},
  journal={Journal of Machine Learning Research},
  volume={23},
  number={132},
  pages={1--109},
  year={2022}
}

@article{kundu2025decomposition,
  title={Decomposition-based intrinsic modeling of shape-constrained functional data},
  author={Kundu, Poorbita and M{\"u}ller, Hans-Georg},
  journal={Electronic Journal of Statistics},
  volume={19},
  number={2},
  pages={4673--4724},
  year={2025},
  publisher={The Institute of Mathematical Statistics and the Bernoulli Society}
}

@article{meyer2008inference,
  author  = {Meyer, Mary C.},
  title   = {Inference Using Shape-Restricted Regression Splines},
  journal = {The Annals of Applied Statistics},
  volume  = {2},
  number  = {3},
  pages   = {1013--1033},
  year    = {2008},
  url     = {http://www.jstor.org/stable/30245118}
}

@article{wu2024shape,
  title={Shape-based functional data analysis},
  author={Wu, Yuexuan and Huang, Chao and Srivastava, Anuj},
  journal={Test},
  volume={33},
  pages={1--47},
  year={2024},
  publisher={Springer}
}

@book{srivastava2016functional,
  title={Functional and shape data analysis},
  author={Srivastava, Anuj and Klassen, Eric P},
  volume={1},
  year={2016},
  publisher={Springer}
}

@article{marron2015functional,
  title={Functional data analysis of amplitude and phase variation},
  author={Marron, James Stephen and Ramsay, James O and Sangalli, Laura M and Srivastava, Anuj},
  journal={Statistical Science},
  pages={468--484},
  year={2015},
  publisher={JSTOR}
}

@article{ramsay1998estimating,
  title={Estimating smooth monotone functions},
  author={Ramsay, James O},
  journal={Journal of the Royal Statistical Society: Series B (Statistical Methodology)},
  volume={60},
  pages={365--375},
  year={1998},
  publisher={Wiley Online Library}
}

@phdthesis{filipini2025tree,
  title={Tree-based models with basis functions},
  author={Filipini dos Santos, Pedro Henrique},
  year={2025}
}

@article{herren2025stochtree,
  title={{StochTree: BART-based modeling in R and Python}},
  author={Herren, Andrew and Hahn, P Richard and Murray, Jared and Carvalho, Carlos},
  journal={arXiv preprint arXiv:2512.12051},
  year={2025}
}

@article{prado2021bayesian,
  title={Bayesian additive regression trees with model trees},
  author={Prado, Estev{\~a}o B and Moral, Rafael A and Parnell, Andrew C},
  journal={Statistics and Computing},
  volume={31},
  pages={20},
  year={2021},
  publisher={Springer}
}

@article{mammen1999smoothing,
  title={Smoothing splines and shape restrictions},
  author={Mammen, Enno and Thomas-Agnan, Christine},
  journal={Scandinavian Journal of Statistics},
  volume={26},
  number={2},
  pages={239--252},
  year={1999},
  publisher={Wiley Online Library}
}

\end{document}